\documentclass[journal,onecolumn,12pt]{IEEEtran}

\usepackage{graphicx}
\usepackage{amsmath}
\usepackage{amssymb}
\usepackage{epic}
\usepackage[left=1in,top=1in,right=1in,bottom=1in]{geometry}

\usepackage{graphics}
\usepackage{epsfig}
\usepackage{latexsym}
\usepackage{cite}

\allowdisplaybreaks

\newcommand{\qed}{\hspace*{\fill} $\Box$ \\}
\newcommand{\bs}{\boldsymbol}

\def\ba{\begin{array}}
\def\ea{\end{array}}
\def\be{\begin{equation}}
\def\ee{\end{equation}}
\def\bea{\begin{eqnarray}}
\def\eea{\end{eqnarray}}
\def\beas{\begin{eqnarray*}}
\def\eeas{\end{eqnarray*}}

\newtheorem{theorem}{Theorem}

\newtheorem{lemma}{Lemma}

\title{Node-Based Optimal Power Control, Routing, and Congestion Control in Wireless Networks}
\author{Yufang Xi and Edmund M. Yeh\\
    Department of Electrical Engineering\\
    Yale University\\
    New Haven, CT 06520, USA\\
    {\tt \{yufang.xi,edmund.yeh\}@yale.edu}}

\begin{document}

\maketitle

\setcounter{footnote}{1} \footnotetext{This research is supported
in part by Army Research Office (ARO) Young Investigator Program
(YIP) grant DAAD19-03-1-0229 and by National Science Foundation
(NSF) grant CCR-0313183.}

\begin{abstract}
We present a unified analytical framework within which power
control, rate allocation, routing, and congestion control for
wireless networks can be optimized in a coherent and integrated
manner. We consider a multi-commodity flow model with an
interference-limited physical-layer scheme in which power control
and routing variables are chosen to minimize the sum of convex link
costs reflecting, for instance, queuing delay. Distributed network
algorithms where joint power control and routing are performed on a
node-by-node basis are presented. We show that with appropriately
chosen parameters, these algorithms iteratively converge to the
global optimum from any initial point with finite cost.  Next, we
study refinements of the algorithms for more accurate link capacity
models, and extend the results to wireless networks where the
physical-layer achievable rate region is given by an arbitrary
convex set, and the link costs are strictly quasiconvex.  Finally,
we demonstrate that congestion control can be seamlessly
incorporated into our framework, so that algorithms developed for
power control and routing can naturally be extended to optimize user
input rates.
\end{abstract}


\section{\label{sec:Introduction}Introduction}

In wireless networks, link capacities are variable quantities
determined by transmission powers, channel fading levels, user
mobility, as well as the underlying coding and modulation schemes.
In view of this, the traditional problems of routing and congestion
control must now be jointly optimized with power control and rate
allocation at the physical layer.  Moreover, the inherent
decentralized nature of wireless networks mandates that distributed
network algorithms requiring limited communication overhead be
developed to implement this joint optimization. In this paper, we
present a unified analytical framework within which power control,
rate allocation, routing, and congestion control for wireless
networks can be optimized in a coherent and integrated manner.  We
then develop a set of distributed network algorithms which
iteratively converge to a jointly optimal operating point. These
algorithms operate on the basis of marginal-cost message exchanges,
and are adaptive to changes in network topology and traffic
patterns.  The algorithms are shown to have superior performance
relative to existing wireless network protocols.


The development of network optimization began with the study of
traffic routing in wireline networks.  Elegant frameworks for
optimal routing within a multi-commodity flow setting are given
in~\cite{paper:CaG74,paper:Gal77}. A distributed routing algorithm
based on gradient projection is developed \cite{paper:Gal77}, where
all nodes iteratively adjust their traffic allocation for each type
of traversing flow.  This algorithm is generalized
in~\cite{paper:BGG84}, where estimates of second derivatives of the
cost function are utilized to improve the convergence rate.

With the advent of variable-rate communications, congestion control
in wireline networks has become an important topic of investigation.
In~\cite{paper:Kel97,paper:KMT98,paper:LL99,book:Sri04}, congestion
control is optimized by maximizing the utilities of contending
sessions with elastic rate demands subject to link capacity
constraints. Distributed algorithms where sources adjust input rates
based on price signal feedback from links are shown to converge to
the optimal operating point. These results have been extended
in~\cite{paper:AA03,paper:WPL03,paper:LS03}, where combined
congestion control and routing (both single-path and multi-path)
algorithms are developed.  The above-mentioned papers generally
consider source routing, where it is assumed that all available
paths to the destinations are known \textit{a priori} at the source
node, which makes the routing decisions.

Wireless networks differ fundamentally from wireline networks in
that link capacities are variable quantities that can be controlled
by adjusting transmission powers.  The power control problem has
been most extensively studied for CDMA wireless networks.  Previous
work at the physical
layer~\cite{paper:FM95,paper:HT99,paper:CS02,paper:JCO01,paper:CS03,paper:BCP00}
has generally focused on developing distributed algorithms to
achieve the optimal trade-off between transmission power levels and
Signal-to-Interference-plus-Noise-Ratios (SINR). More recently,
cross-layer optimization for wireless networks has been investigated
in~\cite{paper:BK04,paper:JXB03,paper:CS03}.  In particular, the
work in~\cite{paper:Chi04} develop distributed algorithms to
accomplish joint optimization of the physical and transport layers
within a CDMA context.

In this work, we present a unified framework in which the power
control, rate allocation, routing, and congestion control
functionalities at the physical, Medium Access Control (MAC),
network, and transport layers of the wireless network can be jointly
optimized. We focus on quasi-static network scenarios where user
traffic statistics and channel conditions vary slowly.  We adopt a
multi-commodity flow model and pose a general problem in which
capacity allocation and routing are jointly optimized to minimize
the sum of convex link costs reflecting, for instance, queuing delay
in the network.   To be specific, we focus initially on an
interference-limited wireless networks where the link capacity is a
concave function of the link SINR.  For these networks, power
control and routing variables are chosen to minimize the total
network cost.  In view of frequent changes in wireless network
topology and node activity, it may not be practical nor even
desirable for sources to obtain full knowledge of all available
paths. We therefore focus on distributed schemes where joint power
control and routing is performed on a {\em node-by-node} basis. Each
node decides on its total transmission power as well as the power
allocation and traffic allocation on its outgoing links based on a
limited number of control messages from other nodes in the network.

We first establish a set of necessary and sufficient conditions for
the joint optimality of a power control and routing configuration.
We then develop a class of node-based scaled gradient projection
algorithms employing first derivative marginal costs which can
iteratively converge to the optimal operating point, without
knowledge of global network topology or traffic patterns.  For rapid
and guaranteed convergence, we develop a new set of upper bounds on
the matrices of second derivatives to scale the direction of
descent. We explicitly demonstrate how the algorithms' parameters
can be determined by individual nodes using limited communication
overhead. The iterative algorithms are rigorously shown to rapidly
converge to the optimal operating point from any initial
configuration with finite cost.


After developing power control and routing algorithms for specific
interference-limited systems, we consider wireless networks with
more general coding/modulation schemes where the physical-layer
achievable rate region is given by an arbitrary convex set. The
necessary and sufficient conditions for the joint optimality of a
capacity allocation and routing configuration are characterized
within this general context.  Under the relaxed requirement that
link cost functions are only strictly quasiconvex, we show that any
operating point satisfying the above conditions is Pareto optimal.

Next, we show that congestion control for users with elastic rate demands can be
seamlessly incorporated into our analytical framework.  We consider maximizing the
aggregate session utility minus the total network cost. It is shown that with the
introduction of virtual overflow links, the problem of jointly optimizing power control,
routing, and congestion control can be made equivalent to a problem involving only power
control and routing in a virtual wireless network.  In this way, the distributed
algorithms previously developed for power control and routing can be naturally extended
to this more general setting.

Finally, we present results from numerical experiments.  The results
confirm the superior performance of the proposed network control
algorithms relative to that of existing wireless network protocols
such as the Ad hoc On Demand Distance Vector (AODV) routing
algorithm~\cite{paper:PR99}.  Our algorithms are shown to converge
rapidly to the optimal operating point.  Moreover, the algorithms
can adaptively chase the shifting optimal operating point in the
presence of slow changes in the network topology and traffic
conditions. Finally, the algorithms exhibit reasonably good
convergence even with delayed and noisy control messages.


The paper is organized as follows.  The basic system model and the
jointly optimal capacity allocation and routing problem formulation
are described in Section~\ref{sec:ProblemFormulation}. In
Section~\ref{sec:OptimizationSpaces}, we specify the jointly optimal
power control and routing problem in node-based form for an
interference-limited wireless network. In
Section~\ref{sec:OptimalityConditions}, the necessary and sufficient
conditions for optimality are presented and proved.  In
Section~\ref{sec:ConvergenceOfAlgorithms}, we present a class of
scaled gradient projection algorithms and characterize the
appropriate algorithm parameters for convergence to the optimum. In
Section~\ref{sec:GeneralizationAndRelaxation}, we develop network
control schemes for more refined link capacity models and derive
optimality results for general convex capacity regions and
quasi-convex cost functions.  Section~\ref{sec:ElasticDemand}
extends the algorithms to incorporate congestion control mechanisms.
Finally, results of relevant numerical experiments are shown in
Section~\ref{sec:Simulation}.


\section{\label{sec:ProblemFormulation}Network Model and Problem Formulation}

\vspace{3mm}\subsection{Network Model, Capacity Region, and Flow Model}

\vspace{3mm}Let the multi-hop wireless network be modelled by a
directed and (strongly) connected graph $\mathcal G = ( {\cal N},
{\cal E} )$, where ${\cal N}$ and ${\cal E}$ are the node and link
sets, respectively. A node $i\in {\cal N}$ represents a wireless
transceiver containing a transmitter with individual power
constraint $\bar P_i$ and a receiver {with additive white Gaussian
noise (AWGN) of power $N_i$}. A link $(i,j) \in {\cal E}$
corresponds to a unidirectional link, which models a radio channel
from node $i$ to $j$.\footnote{We think of ${\cal E}$ as being
predetermined by the communication system setup. For instance, in a
CDMA system, {$(i,j) \in {\cal E}$ if node $j$ knows the spreading
code used by $i$.}} For $(i,j)\in {\cal E}$, let $C_{ij}$ denote its
capacity (in bits/sec). In a wireless network, the value of $C_{ij}$
is variable (we address this issue in depth below).

{A link capacity vector $\boldsymbol C \triangleq ( C_{ij} )_{(i,j)\in {\cal E}}$ is
feasible if it lies in a given \emph{achievable rate region} ${\cal C}\subset
\mathbb{R}_+^{|{\cal E}|}$, which is determined, for example, by the network
coding/decoding scheme and the nodes' transmission powers. In the following, we will
first consider the specific rate region induced by a CDMA-based network model and then
study the more general case of arbitrary convex rate regions in
Section~\ref{sec:GenCapRegion}.}

Consider a collection ${\cal W}$ of communication sessions, each identified by its
source-destination node pair. We adopt a {\em flow model}~\cite{book:BG92} to analyze the
transmission of the sessions' data inside the network.  The flow model is reasonable for
networks where the traffic statistics change slowly over time.\footnote{Such is the case
when each session consists of a large number of independent data streams modelled by
stochastic arrival processes, and no individual process contributes significantly to the
aggregate session rate~\cite{book:BG92}.} As we show, the flow model is particularly
amenable to cost minimization and distributed computation.

For any session $w\in {\cal W}$, let $O(w)$ and $D(w)$ denote the
origin and destination nodes, respectively. Denote session $w$'s
flow rate on link $(i,j)$ by $f_{ij}(w)$. For now, assume the total
incoming rate of session $w$ is a positive constant
$r_w$.\footnote{{Later in Section~\ref{sec:ElasticDemand}, we will
consider elastic sessions with variable incoming rate.}} Thus, we
have the following flow conservation relations. For all $w \in {\cal
W}$,
\begin{equation}\label{eq:FlowConservation}
\begin{array}{ll}
\vspace{1mm} f_{ij}(w) \ge 0, \; & \forall (i,j) \in {\cal E}, \\
\vspace{1mm} \displaystyle\sum_{j \in {\cal O}(i)}{f_{ij}(w)} = r_w \triangleq t_i(w), & i = O(w), \\
\vspace{1mm} f_{ij}(w) = 0, & i = D(w)~\textrm{and}~\forall j \in {\cal O}(i),\\
\vspace{1mm} \displaystyle\sum_{j \in {\cal O}(i)}{f_{ij}(w)} = \sum_{j \in {\cal I}(i)}{f_{ji}(w)} \triangleq t_i(w), & \forall i \ne O(w),D(w), \\
\end{array}
\end{equation}
where ${\cal O}(i) \triangleq \{ j:(i,j) \in {\cal E} \}$ and ${\cal I}(i) \triangleq \{ j:(j,i) \in {\cal E} \}$.
Here, $t_i(w)$ denotes the total incoming rate of session $w$'s traffic at node $i$. 
Finally, the total flow rate on a link is the sum of flow rates of all the sessions using that link:
\begin{equation}\label{eq:LinkRate}
F_{ij} = \sum_{w \in {\cal W}}{f_{ij}(w)}, \; \forall (i,j) \in {\cal E}.
\end{equation}

\vspace{3mm}\subsection{Impact of Traffic Flow and Link Capacities on Network Cost}

\label{sec:FlowCapacityCost}

We assume the network cost is the sum of costs on all the
links.\footnote{If costs also exist at nodes, they can be absorbed
into the costs of the nodes' adjacent links.}  The cost on link
$(i,j)$ is given by a function $D_{ij}(C_{ij}, F_{ij})$ of the
capacity $C_{ij}$ and the total flow rate $F_{ij}$. We assume that
$D_{ij}(C_{ij}, F_{ij})$ is increasing and convex in $F_{ij}$ for
each $C_{ij}$, and decreasing and convex in $C_{ij}$ for each
$F_{ij}$. The link cost function $D_{ij}(C_{ij}, F_{ij})$ can
represent, for instance, the expected delay in the queue served by
link $(i,j)$ with arrival rate $F_{ij}$ and service rate
$C_{ij}$.\footnote{Note that when $C_{ij}$ is fixed, $D_{ij}(C_{ij},
\cdot)$ reduces to the flow-dependent delay function considered in
past literature on optimal routing in wireline networks
\cite{paper:Gal77,paper:BGG84,paper:TsB86}.}  While the monotonicity
of $D_{ij}$ is easy to see, the convexity of $D_{ij}$ in $F_{ij}$
and $C_{ij}$ follows from the fact that the expected queuing delay
increases with the variance of the arrival and/or service
times.\footnote{This phenomenon is captured by the heavy traffic
mean formula for a GI/GI/1 queue with random service time $X$ and
arrival time $A$. The expected waiting time is given by
\[ E[W]\sim \frac{\rho^2 c_x^2 + \rho c_a^2 - \rho(1-\rho)}{2\lambda(1-\rho)}.\] Here,
$\lambda$ denotes the average arrival rate, $\rho=\lambda E[X]$,
$c_x^2 = {\textrm{var}[X]}/{E[X]^2}$ and $c_a^2 =
{\textrm{var}[A]}/{E[A]^2}$.}

\begin{figure}
\begin{center}
\includegraphics[width = 10cm]{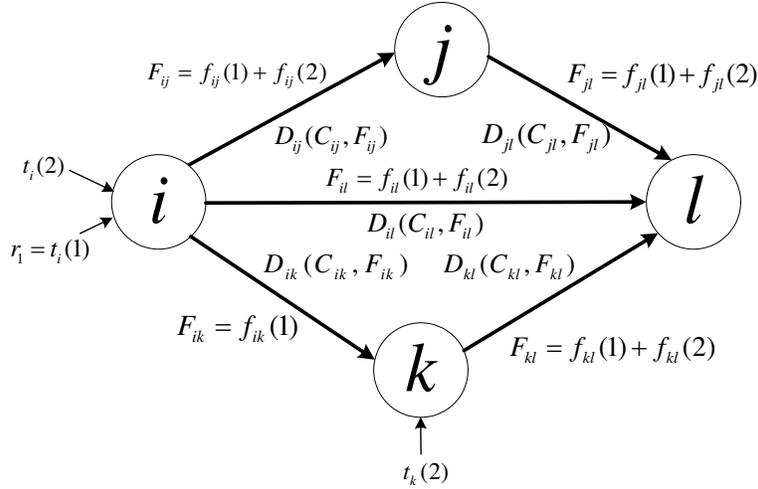}
\caption{Session 1 originates from node $i$ and ends at node $l$. Session 2, originating elsewhere in the network and
destined also for node $l$, enters this part of the network at nodes $i$ and $k$. Node $i$ routes session 1 to $j$,
$k$, and $l$, and routes session 2 to $j$ and $l$. Node $k$ forwards session 2 directly to $l$. These individual flows
make up the total flows on the links. Link costs are determined by the flow rates and capacities.}
\label{fig:FlowCapacity}
\end{center}
\end{figure}

For analytical purposes, $D_{ij}( C_{ij}, F_{ij})$ is further
assumed to be twice continuously differentiable in the region ${\cal
X} = \{(C_{ij}, F_{ij}): 0 \leq F_{ij} < C_{ij}\}$. Moreover, to
implicitly impose the link capacity constraint, we assume $D_{ij}(
C_{ij}, F_{ij})\to\infty$ as $F_{ij}\to C_{ij}^-$ and $D_{ij}(
C_{ij}, F_{ij}) = \infty$ for $F_{ij} \ge C_{ij}$. To summarize, for
all $(i,j) \in {\cal E}$, the cost function $D_{ij}:\mathbb{R}_+
\times \mathbb{R}_+ \mapsto \mathbb{R}_+$ satisfies
\begin{equation}\label{eq:CostFunctionIdentity}
\frac{\partial D_{ij}}{\partial C_{ij}} < 0, \; \frac{\partial D_{ij}}{\partial F_{ij}} >
0, \; \frac{\partial ^2 D_{ij}}{\partial C_{ij} ^2 } \ge 0,  \; \frac{\partial ^2
D_{ij}}{\partial F_{ij} ^2} \ge 0, \quad\textrm{if}~( C_{ij}, F_{ij}) \in {\cal X},
\end{equation}
and $D_{ij}( C_{ij}, F_{ij}) = \infty$ otherwise. As an
example,\footnote{To be precise, an infinitesimal term $\varepsilon$
needs to be added to the numerator, i.e.,
$D_{ij}={(F_{ij}+\varepsilon)}/{(C_{ij}-F_{ij})}$, to make
${\partial D_{ij}}/{\partial C_{ij}} < 0$ for $F_{ij}=0$. }
\begin{equation}\label{eq:ExpectedDelay}
D_{ij}(C_{ij},F_{ij})=\frac{F_{ij}}{C_{ij}-F_{ij}}, \; \textrm{for}
~ 0 \le F_{ij} < C_{ij}
\end{equation}
gives the expected number of packets waiting for or under transmission at link $(i,j)$
under an $M/M/1$ queuing model. Summing over all links, the network cost
$\sum_{(i,j)}D_{ij}(C_{ij},F_{ij})$ gives the average number of packets in the
network.\footnote{By the Kleinrock independence approximation and Jackson's Theorem, the
$M/M/1$ queue is a good approximation for the behavior of individual links when the
system involves Poisson stream arrivals at the entry points, a densely connected network,
and moderate-to-heavy traffic load \cite{book:Kle64,book:BG92}.} {As another example,
$D_{ij} = 1 / (C_{ij}-F_{ij})$ gives the average waiting time of a packet in an M/M/1
queuing model.} The network model and cost functions are illustrated in Figure
\ref{fig:FlowCapacity}.

\subsection{Basic Optimization Problem: Capacity Allocation and Routing}

We now formulate the main Jointly Optimal Capacity allocation and Routing (JOCR) problem,
which involves adjusting $\{f_{ij}(w)\}$ and $\{C_{ij}\}$ jointly to minimize total
network cost as follows:
\begin{eqnarray}
\vspace{1mm}
 {\textrm{minimize}} &~& \sum\limits_{(i,j) \in {\cal E}} {D_{ij}(C_{ij},F_{ij})} \label{eq:JOCR}  \\
\vspace{1mm}
 {\textrm{subject to}}& & \textrm{flow conservation constraints in}~\eqref{eq:FlowConservation}-\eqref{eq:LinkRate}, \nonumber \\ 
\vspace{1mm}
                      & &\boldsymbol C \in {\cal C}. \label{eq:CapacityConstraint}
\end{eqnarray}
The central concern of this paper is the development of distributed algorithms to solve the JOCR problem in useful
network contexts.


\section{\label{sec:OptimizationSpaces}Optimal Distributed Routing and Power Control}

\subsection{Node-Based Routing}

To solve the JOCR problem, we first investigate distributed routing schemes for adapting
link flow rates. In previous literature, there have been extensive discussion of
multi-path \emph{source routing} methods in wireline
networks~\cite{paper:LS03,paper:KST01,paper:WPL03}. In these methods, source nodes are
assumed to have comprehensive information about all available paths through the network
to their destinations. In contrast to wireline networks, however, wireless networks are
characterized by frequent node activity and network topology changes. In these
circumstances, it may not be practical nor even desirable to implement source routing,
which requires source nodes to constantly obtain current path information. We therefore
focus on distributed schemes where routing is performed on a \emph{node-by-node} basis
\cite{paper:Gal77}.  In essence, these schemes distribute routing decisions to all nodes
in the network, rather than concentrating them at source nodes only. As we show, neither
source nodes nor intermediate nodes are required to know the topology of the entire
network. Nodes interact only with their immediate neighbors.

To make distributed adjustment possible, we adopt the {\em routing variables} introduced by
Gallager\cite{paper:Gal77}. They are defined for all $i \in {\cal N}$ and $w \in {\cal W}$ in terms of {\em
link flow fractions} as \be\label{eq:RoutingVariable} \textrm{Routing variables:} \quad \phi_{ij}(w)
\triangleq \displaystyle\frac{f_{ij}(w)}{t_i(w)}, \; j \in {\cal O}(i). \ee

The flow conservation constraints \eqref{eq:FlowConservation} are translated into the
space of routing variables as
\begin{equation}\label{eq:RoutingVariableConstraint}
\begin{array}{ll}
\vspace{1mm} \phi_{ij}(w) \ge 0, \; &\forall j \in {\cal O}(i),
\\
\vspace{1mm} \sum\limits_{j \in {\cal O}(i)}{\phi_{ij}(w)}  = 1, \; &\textrm{if}~i \ne D(w),
\\
\vspace{1mm} \phi_{ij}(w) = 0, \; &\forall j \in {\cal O}(i)~\textrm{if}~i = D(w).
\end{array}
\end{equation}
For node $i$ such that $t_i (w) = 0$, the specific values of $\phi_{ij}(w)$'s are immaterial to the actual flow rates.
They can be assigned arbitrary values satisfying~\eqref{eq:RoutingVariableConstraint}.

The routing variables $(\phi_{ij}(w))_{w \in {\cal W}, (i,j)\in {\cal E} }$ determine the
routing pattern and flow distribution of the sessions. They can be implemented at each
node $i$ using either a deterministic scheme (node $i$ routes $\phi_{ij}(w)$ of its
incoming session-$w$ traffic to neighbor $j$) or a random scheme (node $i$ forwards
session $w$ traffic to $j$ with probability $\phi_{ij}(w)$).  

\subsection{Power Control and Link Capacity}\label{subsec:CapacityModel}

After examining the routing issue, we now address the question of capacity allocation. In
a wireless communication network, given fixed channel conditions, the achievable rate
region ${\cal C}$ is determined by the coding/decoding scheme and transmission powers,
among other factors. To be specific, we focus initially on a wireless network with an
interference-limited physical-layer scheme.

Assume the link capacity $C_{ij}$ is a function $C(SINR_{ij})$ of the
signal-to-interference-plus-noise ratio (SINR) at the receiver of link $(i,j)$, given by
\[
SINR_{ij} = \frac{ G_{ij} P_{ij}}{G_{ij} \sum_{n \ne j} P_{in} + \sum_{m \ne i} G_{mj} \sum_n P_{mn} + N_j},
\]
where $P_{mn}$ is the transmission power on link $(m,n)$, $G_{mj}$ denotes the (constant)
path gain from node $m$ to $j$, $N_j$ is the noise power at node $j$'s receiver.  We
further assume $C(\cdot)$ is strictly increasing, concave, and twice continuously
differentiable.  {For example, in a spread-spectrum CDMA network using (optimal)
single-user decoding, the SINR per symbol is $K \cdot SINR_{ij}$ where $K$ denotes the
processing gain~\cite{book:Vit95}. Since $K$ typically is very large, the
information-theoretic link capacity $\frac{R_s}{2}\log (1 + K \cdot SINR_{ij})$ (in
bits/sec) is well approximated as \be\label{eq:HighSINRCapacity} C_{ij} \approx
\frac{R_s}{2}\log ( K \cdot SINR_{ij}), \ee where $R_s$ is the (fixed) symbol rate of the
CDMA sequence. As another example, if messages are modulated on CDMA symbols using M-QAM,
and the error probability is required to be less than or equal to $\bar P_e$, then the
maximum data rate under the same high-SINR assumption is given by~\cite{book:TV04}
\be\label{eq:MQAMCapacity} C_{ij} = R_s \log\left(\frac{K\cdot SINR_{ij}}{2
\left[Q^{-1}(\bar P_e)\right]^2} \right), \ee where $Q(\cdot)$ is the complementary
distribution function of a normal random variable.}

Assume that every node is subject to an individual power constraint
\be\label{eq:TotalPowerConstraint} P_i \triangleq \sum_{j \in {\cal O}(i)} P_{ij} \le
\bar P_i.\ee Denote the set of feasible power vectors $\bs P \triangleq
(P_{ij})_{(i,j)\in{\cal E}}$ by $\Pi$.

We now note that the objective function in~\eqref{eq:JOCR},
$\sum_{(i,j)} D_{ij}(C_{ij}(\bs P), F_{ij})$, is convex in the flow
variables $(F_{ij})$. It is convex in $\bs P$ if every $C_{ij}$ is
concave in $\bs P$. Unfortunately, given that $C_{ij} =
C(SINR_{ij})$ is strictly increasing, $\nabla^2 C_{ij}(\bs P)$
cannot be negative definite. However, it is observed
in~\cite{paper:HBH05} that if \be\label{eq:ConcaveCapacity} C''(x)
\cdot x + C'(x) \le 0, \quad \forall x \ge 0,\ee then with a change
of variables $S_{mn} = \ln P_{mn}$~\cite{paper:Chi04}, $C_{ij}$ is
concave in $\bs S \triangleq (S_{mn})_{(m,n)\in{\cal E}}$.  From
this, it can be verified that the objective function
in~\eqref{eq:JOCR} is convex in $\bs S $. In the following, we
assume $C(\cdot)$ satisfies \eqref{eq:ConcaveCapacity}.  Note that
this is true for the capacity functions of the CDMA and M-QAM
examples above.  For brevity, we will sometimes denote $SINR_{ij}$
by $x_{ij}$. We will also make use of the log-power variables $\bs
S$ (i.e., power measured in dB), which belong to the feasible set
$\Pi_{\bs S} = \{ \bs S \in \mathbb{R}^{|{\cal E}|}: \sum_{j \in
{\cal O}(i)} e^{S_{ij}} \le \bar P_i, \forall i \in {\cal N} \}$.

As in the case of the routing variables $\phi_{ij}(w)$, it is convenient to express the transmission power
$P_{ij}$ on link $(i,j)$ in terms of the \emph{power control} and \emph{power allocation} variables as
follows:
\begin{eqnarray}
\vspace{1mm} \textrm{Power allocation variables:} &\;& \eta _{ij} \triangleq \displaystyle\frac{P_{ij}}{P_i}, \;
(i,j) \in {\cal E}, \label{eq:PowerAllocVariable} \\
\vspace{1mm} \textrm{Power control variables:} & & \gamma_i \triangleq \displaystyle\frac{\ln P_i}{\ln \bar P_i}, \; i
\in {\cal N}. \label{eq:PowerCtrlVariable}
\end{eqnarray}


With appropriate scaling, we can always let all $\bar P_i
> 1$  so that the constraints for $\eta_{ij}$ and $\gamma_i$ can be written as follows:
\begin{equation}\label{eq:PowerVariableConstraint}
\vspace{1mm} \eta _{ij} \ge 0, \; \forall (i,j) \in {\cal E}, \quad\displaystyle\sum_{j \in {\cal O}(i)}{\eta_{ij}} =
1, \; \gamma_i \le 1, \; \forall i \in {\cal N}.
\end{equation}

\subsection{Distributed Optimization Problem: Power Control and Routing}

With definitions \eqref{eq:RoutingVariable}, \eqref{eq:PowerAllocVariable}, and
\eqref{eq:PowerCtrlVariable}, the JOCR problem in \eqref{eq:JOCR} can be expressed in
node-based form. We call this the Jointly Optimal Power Control and Routing (JOPR)
problem:
\begin{eqnarray}
\vspace{1mm}{\textrm{minimize}} &\; & \sum\limits_{(i,j) \in {\cal E}} {D_{ij}(C_{ij},F_{ij})}\label{eq:JOPR} \\
\vspace{1mm}{\textrm{subject to}} & & \eqref{eq:RoutingVariableConstraint}, \eqref{eq:PowerVariableConstraint}, 
\label{eq:PCVarConstraint}
\end{eqnarray}
where link flow rates and capacities are determined by\footnote{Notice that in general,
$C_{ij}$ should be upper bounded by the RHS of \eqref{eq:CapacityFormulaApprox}. However,
since cost function $D_{ij}(\cdot, F_{ij})$ is decreasing in $C_{ij}$, any solution of
problem \eqref{eq:JOCR} must allocate a vector of link capacities lying on the boundary
of ${\cal C}$. Therefore, without loss of optimality, we assume equality in
\eqref{eq:CapacityFormulaApprox}.}
\begin{eqnarray}
\vspace{1mm} F_{ij} &=& \displaystyle\sum_{w \in {\cal W}}{t_i(w)
\cdot \phi_{ij}(w)}, \quad \forall (i,j) \in {\cal E}, \label{eq:LinkFlowRate} \\
\vspace{1mm} t_i(w) &=& \left\{ \begin{array}{ll}
                            \vspace{1mm} r_w, \; &i =
                            O(w) \\
                            \displaystyle\sum_{j \in {\cal I}(i)}{t_j(w) \cdot \phi_{ji}(w)},
                            &\forall i \neq O(w)
                            \end{array} \right. , \label{eq:FlowBalance} \\
\vspace{1mm} C_{ij} &=& C\left(\displaystyle\frac{G_{ij} (\bar P_i)^{\gamma_i}
\eta_{ij}}{ G_{ij} (\bar P_i)^{\gamma_i} \displaystyle\sum_{k \ne j}{\eta_{ik}} + \sum_{m
\ne i}{G_{mj} (\bar P_m)^{\gamma_m}} + N_j }\right), \quad \forall (i,j) \in {\cal E}.
\label{eq:CapacityFormulaApprox}
\end{eqnarray}

\vspace{3mm}
\section{\label{sec:OptimalityConditions}Conditions for Optimality}

To specify the optimality conditions for the JOPR problem in \eqref{eq:JOPR}, it is necessary to compute the
cost gradients with respect to the routing variables, the power allocation variables, and the power control
variables, respectively.  For the routing variables, the gradients are given in~\cite{paper:Gal77} as
\begin{equation}\label{eq:RoutingVarGrad}
\frac{\partial D}{\partial \phi_{ij}(w)} = t_i(w) \cdot \delta\phi_{ij}(w), \; \forall j \in {\cal O}(i),
\end{equation}
where the \emph{marginal routing cost} is
\begin{equation}\label{eq:MarginalFlowCost}
\delta\phi_{ij}(w) \triangleq \frac{\partial D_{ij}}{\partial F_{ij}} + \frac{\partial D}{\partial r_j(w)}.
\end{equation}
Here, $\frac{\partial D}{\partial r_j(w)}$ stands for the marginal cost due to a unit increment of session $w$'s input
traffic at $j$.  It is computed recursively by~\cite{paper:Gal77}
\begin{eqnarray}
\displaystyle\frac{\partial D}{\partial r_j(w)} &=& 0, \quad \textrm{if}~j = D(w), \label{eq:NodeMarginalRoutingCost1}\\
\displaystyle\frac{\partial D}{\partial r_i(w)} &=& \sum\limits_{j \in {\cal
O}(i)}{\phi_{ij}(w)\left[\frac{\partial D_{ij}}{\partial F_{ij}} + \frac{\partial D}{\partial r_j(w)}\right]}
= \sum\limits_{j \in {\cal O}(i)}{\phi_{ij}(w) \cdot \delta\phi_{ij}(w)}, \; \forall i \ne D(w).
\label{eq:NodeMarginalRoutingCost2}
\end{eqnarray}

We now compute the gradients with respect to the power allocation and power control variables:
\begin{equation}\label{eq:PowerAllocationGrad}
\frac{\partial D}{\partial \eta_{ij}} = P_i \left[ -\sum_{(m,n)} \frac{\partial D_{mn}}{\partial C_{mn}}
\frac{C_{mn}' G_{mn} G_{in} P_{mn}}{IN_{mn}^2} + \delta\eta_{ij} \right],
\end{equation}
where $C_{mn}'$ is short-hand notation for $d C(x_{mn}) / d x_{mn}$. Here, the
\emph{marginal power allocation cost} is
\begin{equation}\label{eq:MarginalPowerAlloc}
\delta\eta_{ij} \triangleq \frac{\partial D_{ij}}{\partial C_{ij}} \frac{C_{ij}' G_{ij}}{IN_{ij}}(1 + SINR_{ij}).
\end{equation}
Finally, the derivatives with respect to the power control variables are given by
\begin{equation}\label{eq:PowerCtrlGrad}
\frac{\partial D}{\partial \gamma_{i}} = \bar S_i \cdot \delta\gamma_i,
\end{equation}
where the \emph{marginal power control cost} is \be\label{eq:MarginalPowerCtrl}
\delta\gamma_i \triangleq P_i \left[ -\sum_{(m,n)} \frac{\partial D_{mn}}{\partial
C_{mn}} \frac{C_{mn}' G_{mn} G_{in} P_{mn}}{IN_{mn}^2} + \sum_{j \in {\cal O}(i)}
\delta\eta_{ij} \cdot \eta_{ij} \right]. \ee

The term $IN_{ij}$ appearing in \eqref{eq:PowerAllocationGrad}, \eqref{eq:MarginalPowerAlloc} and
\eqref{eq:MarginalPowerCtrl} is short-hand notation for the overall interference-plus-noise power at the
receiver of link $(i,j)$:
\[
IN_{ij} = G_{ij} \displaystyle\sum_{k \ne j} P_{ik} + \sum\limits_{m \ne i} G_{mj} \sum_{k \in {\cal O}(m)}
P_{mk} + N_j.
\]

We will present the methods for providing nodes with the above marginal costs
$\delta\phi_{ij}(w)$, $\delta\eta_{ij}$ and $\delta\gamma_i$, along with the description
of distributed routing and power adjustment algorithms, in Section
\ref{sec:ConvergenceOfAlgorithms}.


Given the marginal costs $\delta\phi_{ij}(w)$, $\delta\eta_{ij}$, and $\delta\gamma_i$,
each node can check whether optimality is achieved by verifying the conditions stated in
the following theorem, which generalizes Theorem~2 of Gallager~\cite{paper:Gal77} to the
wireless setting.

\vspace{0.1in}\begin{theorem}\label{thm:OptimalityCondition}  Assume
the link cost functions $D_{ij}(C_{ij},F_{ij})$ satisfy the
conditions in~\eqref{eq:CostFunctionIdentity}.  For a feasible set
of routing and transmission power allocations $\{\phi_{ij}(w)\}_{w
\in {\cal W}, (i,j) \in {\cal E} }$, $\{\eta_{ij}\}_{(i,j) \in {\cal
E}}$ and $\{\gamma_i\}_{i \in {\cal N}}$ to be the solution of the
JOPR problem in \eqref{eq:JOPR}, the following conditions are
necessary. For all $w \in {\cal W}$ and $i \neq
 D(w) $ with $t_i(w) > 0$, there exists a constant $\lambda_i(w)$ such that
\begin{eqnarray}
\vspace{1mm} \delta\phi_{ij}(w) = \lambda_i(w), &\;& \textrm{if}~\phi_{ij}(w) > 0 \label{eq:RoutingOptCond1},
\\
\vspace{1mm} \delta\phi_{ij}(w) \ge \lambda_i(w), &\;& \textrm{if}~\phi_{ij}(w) = 0
\label{eq:RoutingOptCond2}.
\end{eqnarray}
For all $i \in {\cal N}$, all $\eta_{ij} > 0$, and there exists a constant $\nu_i$ such
that
\begin{eqnarray}
\vspace{1mm} \delta\eta_{ij} = \nu_i, &\;& \forall j \in {\cal O}(i), \label{eq:PowerOptCond1}
\\
\vspace{1mm} \frac{\delta\gamma_i}{P_i} = 0, &\;& \textrm{if}~\gamma_i < 1, \label{eq:PowerOptCond2}
\\
\vspace{1mm} \frac{\delta\gamma_i}{P_i} \le 0, &\;& \textrm{if}~\gamma_i = 1. \label{eq:PowerOptCond3}
\end{eqnarray}
If the link cost functions $D_{ij}(C_{ij},F_{ij})$ are also jointly
convex in $(C_{ij},F_{ij})$, then these conditions are sufficient
for optimality if
\eqref{eq:RoutingOptCond1}-\eqref{eq:RoutingOptCond2} hold at every
$i \neq D(w)$ for all $w \in {\cal W}$, whether $t_i(w) > 0$ or not.
\end{theorem}\vspace{0.1in}

Note that because $D_{ij}(C_{ij},F_{ij})$ is defined to be infinite
when $C_{ij}=0$ (cf.~Section~\ref{sec:FlowCapacityCost}), we must
have $\eta_{ij}>0$ for all $(i,j)\in{\cal E}$ at the optimum.
Furthermore, note that the sufficiency part of
Theorem~\ref{thm:OptimalityCondition} requires the cost function
$D_{ij}(C_{ij},F_{ij})$ to be jointly convex in $(C_{ij},F_{ij})$.
This is true for the cost function $D_{ij} = 1/(C_{ij}-F_{ij})$ for
$0 \le F_{ij} < C_{ij}$, but not true for the cost function $D_{ij}
= F_{ij}/(C_{ij}-F_{ij})$.   To deal with the latter case, we will
establish the conditions for a Pareto optimal operating point for
strictly quasiconvex cost functions in
Section~\ref{sec:QuasiConvex}.

Before presenting the proof of Theorem
\ref{thm:OptimalityCondition}, we point out a useful identity.
\vspace{0.1in}
\begin{lemma}\label{lma:LinkNodeRateRelation}
With node-based marginal routing costs defined as in \eqref{eq:NodeMarginalRoutingCost1} and
\eqref{eq:NodeMarginalRoutingCost2}, we have
\begin{equation}\label{eq:LinkNodeRateRelation}
\sum_{(i,j) \in {\cal E}}{\frac{\partial D_{ij}}{\partial F_{ij}}(C_{ij},F_{ij}) \cdot F_{ij}} = \sum_{w \in {\cal
W}}{\frac{\partial D}{\partial r_{O(w)}(w)} \cdot r_w}.
\end{equation}
\end{lemma}\vspace{0.1in}

The proof of the lemma requires only algebraic manipulations. It can be found in
Appendix~\ref{app:LinkNodeRateRelation}.\vspace{3mm}

\textit{Proof of Theorem \ref{thm:OptimalityCondition}:} To prove the necessity of
\eqref{eq:RoutingOptCond1}-\eqref{eq:RoutingOptCond2}, suppose it is violated for some $w$ at some node $i
\ne D(w)$ such that $t_i(w) > 0$. By \eqref{eq:RoutingVarGrad}, there exists link $(i,j)$ such that
$f_{ij}(w) = t_i(w) \phi_{ij}(w) > 0$ and
\[
\displaystyle{\frac{\partial D}{\partial \phi_{ij}(w)} > \min_{l \in {\cal O}(i)} \frac{\partial D}{\partial
\phi_{il}(w)}}.
\]
Then by shifting a tiny portion of flow of session $w$ from link $(i,j)$ to a link having
minimal marginal cost, i.e. any link $(i,k)$ such that $k = \arg\min_{l \in {\cal O}(i)}
\frac{\partial D}{\partial \phi_{il}(w)}$, the total cost is decreased. Thus
$\{\phi_{ij}(w)\}$ cannot be optimal. The necessity of conditions
\eqref{eq:PowerOptCond1}-\eqref{eq:PowerOptCond3} can be verified in the same way by
making use of \eqref{eq:PowerAllocationGrad} and \eqref{eq:PowerCtrlGrad}. \vspace{0.1in}

To show the sufficiency statement, assume $\{\phi_{ij}^*(w)\}_{w \in {\cal W},(i,j) \in {\cal E}}$,
$\{\eta_{ij}^*\}_{(i,j) \in {\cal E}}$ and $\{\gamma_i^*\}_{i \in {\cal N}}$ is a set of valid routing and power
variables that satisfy \eqref{eq:RoutingOptCond1}-\eqref{eq:PowerOptCond3}. Let $\{\phi_{ij}^1(w)\}_{w \in {\cal
W},(i,j) \in {\cal E}}$, $\{\eta_{ij}^1\}_{(i,j) \in {\cal E}}$ and $\{\gamma_i^1\}_{i \in {\cal N}}$ be any other set
of feasible routing and power variables. Denote the resulting link flow rates, link capacities and log-powers under
these two schemes by $\{F_{ij}^*\}$, $\{C_{ij}^*\}$, $\{S_{ij}^*\}$ and $\{F_{ij}^1\}$, $\{C_{ij}^1\}$, $\{S_{ij}^1\}$,
respectively. Using the convexity of cost functions and summing over all $(i,j) \in {\cal E}$, we have
\begin{eqnarray}
&&\displaystyle{\sum_{(i,j) \in {\cal E}}{D_{ij}(C_{ij}^1,F_{ij}^1) - D_{ij}(C_{ij}^*,F_{ij}^*)}} \nonumber \\
&\ge& \displaystyle{\sum_{(i,j) \in {\cal E}}{\frac{\partial D_{ij}}{\partial F_{ij}}(C_{ij}^*,F_{ij}^*) \cdot
(F_{ij}^1 - F_{ij}^*)} + \sum_{(i,j) \in {\cal E} }{\frac{\partial D_{ij}}{\partial C_{ij}}(C_{ij}^*,F_{ij}^*) \cdot
(C_{ij}^1 - C_{ij}^*)} }. \label{eq:ObjectiveDiff}
\end{eqnarray}
We show that the two summations on the RHS of \eqref{eq:ObjectiveDiff} are both
non-negative, thus establishing the superiority of $\{\phi_{ij}^*(w)\}$,
$\{\eta_{ij}^*\}$ and $\{\gamma_i^*\}$. We analyze the first summation as follows:
\begin{eqnarray*}
\vspace{1mm} && \displaystyle{\sum_{(i,j) \in {\cal E}}{\frac{\partial D_{ij}}{\partial
F_{ij}}(C_{ij}^*,F_{ij}^*) \cdot
(F_{ij}^1 - F_{ij}^*)}} \\
&\mathop=\limits^{(a)}& \displaystyle{\sum_{(i,j) \in {\cal E}}{\frac{\partial D_{ij}}{\partial F_{ij}}(C_{ij}^*,F_{ij}^*) \cdot F_{ij}^1 } - \sum_{w \in {\cal W}}{\frac{\partial D^*}{\partial r_{O(w)}(w)} \cdot r_w }}  \\
&\mathop=\limits^{(b)}& \displaystyle{\sum_{w \in {\cal W}}{\left[\sum_{(i,j) \in {\cal E}}{\frac{\partial D_{ij}}{\partial F_{ij}}(C_{ij}^*,F_{ij}^*) \cdot t_i^1(w) \phi_{ij}^1(w)}\right]} - \sum_{\substack{w \in {\cal W} \\ i = O(w)}}{\frac{\partial D^*}{\partial r_i(w)} \cdot t_i^1(w)}}\\
& &- \displaystyle{\sum_{w \in {\cal W}}{\left\{\sum_{j \ne O(w),D(w)}{\frac{\partial D^*}{\partial r_j(w)}\left[t_j^1(w) - \sum_{i \ne D(w)}{t_i^1(w) \phi_{ij}^1(w)}\right]}\right\}}}  \\
&\mathop=\limits^{(c)} &\displaystyle{\sum_{w \in {\cal W}}{\left\{\sum_{i \ne D(w)}{t_i^1(w)\left(\sum_{j \in {\cal O}(i)}{\phi_{ij}^1(w)\left[\frac{\partial D_{ij}}{\partial F_{ij}}(C_{ij}^*,F_{ij}^*) + \frac{\partial D^*}{\partial r_j(w)}\right]} - \frac{\partial D^*}{\partial r_i(w)}\right)}\right\}}}  \\
&\mathop=\limits^{(d)} &\displaystyle{\sum_{w \in {\cal W}}{\left\{\sum_{i \ne D(w)}{t_i^1(w)\left[\sum_{j
\in {\cal O}(i)}{\phi_{ij}^1(w) \delta\phi_{ij}^*(w)} - \min_{j \in {\cal
O}(i)}{\delta\phi_{ij}^*(w)}\right]}\right\}}} \mathop\ge\limits^{(e)} 0
\end{eqnarray*}
The first equation results from Lemma \ref{lma:LinkNodeRateRelation}. To obtain (b), we first use the definition of
$F_{ij}^1$ in \eqref{eq:LinkFlowRate} and the fact that $t_i^1(w) = r_w$, $\forall w \in {\cal W}$ and $i = O(w)$. We
then append the zero terms (cf.~\eqref{eq:FlowBalance})
\[\sum_{j \ne O(w),D(w)}{\frac{\partial D^*}{\partial r_j(w)}\left[t_j^1(w) - \sum_{i \ne D(w)}{t_i^1(w)
\phi_{ij}^1(w)}\right]},\]for all $w \in {\cal W}$. By rearranging terms on the RHS of (b), we get equation (c). The
optimality conditions \eqref{eq:RoutingOptCond1}-\eqref{eq:RoutingOptCond2} are translated into equation (d), which
immediately results in inequality (e).

Next, we examine the second summation in \eqref{eq:ObjectiveDiff}. Recalling the concavity of $C_{ij}$ in
terms of $(S_{mn})$ and noticing that $\frac{\partial D_{ij}}{\partial C_{ij}} < 0$, we can bound the second
summation by
\begin{equation}\label{eq:ObjectiveDiffInPower}
\sum_{(i,j) \in {\cal E} }{\frac{\partial D_{ij}}{\partial C_{ij}}(C_{ij}^*,F_{ij}^*) \cdot (C_{ij}^1 -
C_{ij}^*)} \ge \sum_{(i,j) \in {\cal E}}{\frac{\partial D_{ij}^*}{\partial C_{ij}}\sum_{(m,n) \in {\cal
E}}{\frac{\partial C_{ij}^*}{\partial S_{mn}}(S_{mn}^1 - S_{mn}^*)}},
\end{equation}
where $\frac{\partial D_{ij}}{\partial C_{ij}}(C_{ij}^*,F_{ij}^*)$ is abbreviated as $\frac{\partial
D_{ij}^*}{\partial C_{ij}}$ and $\frac{\partial C_{ij}}{\partial S_{mn}}(\boldsymbol S^*)$ is abbreviated as
$\frac{\partial C_{ij}^*}{\partial S_{mn}}$. Differentiating $C_{mn}(\bs S)$ with respect to each of its
variables, we have
\begin{equation}\label{eq:CapacityDerivInS}
\frac{\partial C_{mn}}{\partial S_{ij}} = \left\{\begin{array}{ll}
\vspace{1mm} C'_{mn} \cdot x_{mn}, \; &\textrm{if}~(i,j) = (m,n), \\
\vspace{1mm} -C'_{mn} \cdot x_{mn} \displaystyle\frac{G_{in} P_{ij}}{IN_{mn}}, &\textrm{otherwise},
\end{array} \right.
\end{equation}
where $x_{mn}$ denotes $SINR_{mn}$. We further transform and bound the RHS of \eqref{eq:ObjectiveDiffInPower} as \beas
&&\displaystyle{\sum_{(i,j) \in {\cal E}}{\frac{\partial D_{ij}}{\partial C_{ij}^*}\sum_{(m,n) \in {\cal
E}}{\frac{\partial C_{ij}}{\partial S_{mn}^*} \cdot (S_{mn}^1 - S_{mn}^*)}}}
\\
&\stackrel{(a)}{=}& \sum_{(i,j) \in {\cal E}} \left[ - \sum_{(m,n) \in {\cal E}} \frac{\partial
D_{mn}}{\partial C_{mn}^*}  (C^*_{mn})'  x_{mn}^*  \frac{G_{in}}{IN_{mn}^*} + \nu_i^*\right] P_{ij}^*
 \ln\frac{P_{ij}^1}{P_{ij}^*}
\\
&\stackrel{(b)}{=}& \sum_{(i,j) \in {\cal E}} \frac{\delta\gamma_i^*}{P_i^*} \cdot P_{ij}^*
 \ln\frac{P_{ij}^1}{P_{ij}^*}
\\
&\stackrel{(c)}{\ge}&\displaystyle{\sum_{(i,j) \in {\cal E}}\frac{\delta\gamma_i^*}{P_i^*} \cdot P_{ij}^*
\left(\frac{P_{ij}^1}{P_{ij}^*} - 1\right)}
\\
&\stackrel{(d)}{=}&\displaystyle{\sum_{i \in {\cal N}}\frac{\delta\gamma_i^*}{P_i^*} \cdot (P_i^1 - P_i^*)}
\stackrel{(e)}{\ge} 0. \eeas Here, equality (a) follows from the definition of $\{\delta\eta_{ij}\}$ and the optimality
condition~\eqref{eq:PowerOptCond1}. Using the definition of $\{\delta\gamma_i\}$, we obtain equality (b). By the
conditions~\eqref{eq:PowerOptCond2}-\eqref{eq:PowerOptCond3}, $\delta\gamma_i^* / P_i^* \le 0$. This, together with the
fact that $\ln x \le x - 1,~\forall x \ge 0$, yields inequality (c). Summing over all $j \in {\cal O}(i)$ for each $i
\in {\cal N}$, we obtain (d). The last inequality (e) is implied by
conditions~\eqref{eq:PowerOptCond2}-\eqref{eq:PowerOptCond3} as well.

We have shown that $\sum_{(i,j) \in {\cal E}}{D_{ij}(C_{ij}^1,F_{ij}^1) -
D_{ij}(C_{ij}^*,F_{ij}^*)} \ge 0$ for any $\{\phi_{ij}^1(w)\}$, $\{\eta_{ij}^1\}$ and
$\{\gamma_i^1\}$. Therefore, $\{\phi_{ij}^*(w)\}$, $\{\eta_{ij}^*\}$ and $\{\gamma_i^*\}$
must be an optimal solution. \qed


\vspace{3mm}

\section{\label{sec:ConvergenceOfAlgorithms}Node-Based Network Algorithms}

After obtaining the optimality conditions, we come to the question
of how individual nodes can adjust their local optimization
variables to achieve a globally optimal configuration.  In this
section, we design a set of algorithms that update the nodes'
routing variables, power allocation variables, and power control
variables in a distributed manner, so as to asymptotically converge
to the optimum.

Since the JOPR problem in~\eqref{eq:JOPR} involves the minimization
of a convex objective over convex regions, the class of {\em
gradient projection} algorithms is appropriate for providing a
distributed solution.  An iteration of the gradient projection
method involves making a small update in a direction (typically
opposite of the direction of the gradient) which reduces the network
cost.  Whenever an update leads to a point outside the feasible set,
the point is projected back into the feasible set~\cite{book:Ber99}.
The gradient projection approach was adopted by Gallager for
distributed optimal routing in wireline networks~\cite{paper:Gal77}.
The algorithm in~\cite{paper:Gal77}, although guaranteed to
converge, has a slow rate of convergence due in part to very small
stepsizes. To improve the convergence rate of the gradient
projection algorithms, it is generally necessary to scale the
descent direction using, for instance, second derivatives of the
objective function. In the latter case, the scaled gradient
projection algorithm becomes a version of the projected Newton
algorithm, which is known to enjoy super-linear convergence rates
when the initial point is close to the optimum~\cite{book:Ber99}. In
the current network setting, however, the inherent large
dimensionality and the need for distributed computation preclude
exact calculation of the Hessian required for the Newton algorithm.
Motivated by these considerations, Bertsekas et
al.~\cite{paper:BGG84} developed distributed optimal routing schemes
for wireline networks where diagonal approximations to the Hessian
are used to scale the descent direction. Although the algorithm
in~\cite{paper:BGG84} represents a significant step forward, it
suffers from two major problems.  First, the algorithm
in~\cite{paper:BGG84} is not guaranteed to converge if the initial
point is too far from the optimum. Second, substantial communication
overhead is still required to compute the scaling matrices in a
distributed fashion~\cite{paper:BGG84}.

In this section, we develop a set of scaled gradient projection
algorithms which update the nodes' routing, power allocation, and
power control variables in a distributed manner for a wireless
network. Network protocols which allow for the information exchange
necessary to implement these algorithms are specified.  We develop a
new technique for selecting the scaling matrices for the routing,
power allocation, and power control algorithms based on upper bounds
on the corresponding Hessian matrices.  We show that the resulting
algorithms are guaranteed to converge rapidly to the optimum point
from any initial condition with finite cost. Moreover, we show that
convergence can take place with limited control overhead and
distributed implementation.  In particular, the routing algorithm
exhibits faster convergence than its counterpart
in~\cite{paper:Gal77} and requires less communication overhead than
its counterpart in~\cite{paper:BGG84}.

\subsection{Routing Algorithm (RT)} \label{subsec:RT}

We will develop a suite of algorithms that iteratively adjust a
node's routing, power allocation, and power control variables,
respectively.  First, we present the routing algorithm.

The routing algorithm allows each node to update its routing
variables for all traversing sessions.  We design an algorithm in
the general scaled gradient projection form studied
in~\cite{paper:BGG84}, which contains the algorithm of
Gallager~\cite{paper:Gal77} as a special case.  The scaling matrices
in our routing algorithm, however, are different from those
in~\cite{paper:BGG84}.  We develop a new technique of upper bounding
the relevant Hessians which leads to larger stepsizes, and therefore
faster convergence, than those proposed in~\cite{paper:Gal77}.
Moreover, in contrast to~\cite{paper:BGG84}, our technique
guarantees convergence from any initial condition with finite cost,
and requires less computation and communication overhead to
implement.

\subsubsection{Routing Algorithms of Gallager, Bertsekas, and Gafni~\cite{paper:Gal77,paper:BGG84}}

In order to establish the setting, we first review the (wireline)
routing algorithms of Gallager, Bertsekas, and
Gafni~\cite{paper:Gal77,paper:BGG84}. Consider node $i \ne D(w)$. At
the $k$th iteration, the routing algorithm RT updates the current
routing configuration $\boldsymbol\phi_i^k(w) \triangleq
(\phi_{ij}^k(w))_{j \in {\cal O}(i)}$ by
\begin{equation}\label{eq:GeneralRT}
\boldsymbol{\phi}_i^{k + 1}(w) = RT(\boldsymbol\phi_i^k(w)),
\end{equation}
where the update is determined by the following scaled gradient projection:
\begin{equation}\label{eq:GeneralRTGradProjUpdate}
\boldsymbol\phi_i^{k + 1}(w) = \left[\boldsymbol\phi_i^k(w) - (M_i^k(w))^{-1} \cdot
\delta\boldsymbol\phi_i^k(w)\right]_{M_i^k(w)}^+.
\end{equation}
Here, $\delta\boldsymbol\phi_i^k(w) \triangleq
(\delta\phi_{ij}^k(w)_{j \in {\cal O}(i)} $. The matrix $M_i^k(w)$,
which scales the descent direction for good convergence properties,
is symmetric and positive definite. We will discuss how to choose
$M_i^k(w)$ in a moment. The operator $[\cdot]_{M_i^k(w)}^+$ denotes
projection on the feasible set relative to the norm induced by
matrix $M_i^k(w)$. This is given by
\[
[\tilde{\boldsymbol\phi}_i(w)]_{M_i^k(w)}^+ = \mathop{\arg\min}_{\boldsymbol\phi_i(w) \in
{\cal F}_i^k(w)} \langle\boldsymbol\phi_i(w) - \tilde{\boldsymbol\phi}_i(w),
M_i^k(w)(\boldsymbol\phi_i(w) - \tilde{\boldsymbol\phi}_i(w))\rangle,
\]
where $\langle \cdot \rangle$ denotes the standard Euclidean inner product, and the
minimization is taken over simplex
\[
{\cal F}_i^k(w) = \left\{\boldsymbol\phi_i(w):~\boldsymbol\phi_i(w) \ge \mathbf
0,~\phi_{ij}(w) = 0, ~\forall j \in {\cal B}^k_i(w)~\textrm{and}~\sum\limits_{j \in {\cal
O}(i)}{\phi_{ij}(w)} = 1\right\}.
\]
Here, ${\cal B}^k_i(w)$ represents the set of \emph{blocked nodes}
of $i$ relative to session $w$. This device was invented
in~\cite{paper:Gal77,paper:BGG84} for preventing loops in the
routing pattern of any session.  It contains the neighbors of $i$ to
which $i$ cannot route session-$w$ traffic~. We will discuss ${\cal
B}^k_i(w)$ in more details later.  With straightforward
manipulation, one can show~\cite{paper:BGG84} that the projection
$\boldsymbol\phi_i^{k+1}(w)$ is a solution to
\begin{equation}\label{eq:GeneralRTUpdate}
\begin{array}{ll}
\vspace{1mm} \textrm{minimize} \; &\delta\boldsymbol\phi_i^k(w)' \cdot
\left(\boldsymbol\phi_i(w) - \boldsymbol\phi_i^k(w)\right) +
\displaystyle{\left(\boldsymbol\phi_i(w) - \boldsymbol\phi_i^k(w)\right)' \cdot
\frac{M_i^k(w)}{2 } \cdot \left(\boldsymbol\phi_i(w) - \boldsymbol\phi_i^k(w)\right)}
\\
\vspace{1mm} \textrm{subject to} &\boldsymbol\phi_i(w) \in {\cal F}_i^k(w).
\end{array}\end{equation}
In the following, we use~\eqref{eq:GeneralRTUpdate} to represent the scaled projection
algorithm {and refer to it specifically as the general routing algorithm, or GRT}.

The routing algorithm requires the following two supplementary
mechanisms which coordinate the necessary message exchange and the
suppression of loopy routes in the
network~\cite{paper:Gal77,paper:BGG84}.

\vspace{3mm} {\bf Message Exchange Protocol}:
In order for node $i$ to evaluate the terms $\delta\phi_{ij}(w)$ in
\eqref{eq:MarginalFlowCost}, it needs to collect local measures
${\partial D_{ij}}/{\partial F_{ij}}$ as well as reports of marginal
costs ${\partial D}/{\partial r_j(w)}$ {from its neighbors $j$ to
which it forwards session-$w$ traffic.} Moreover, node $i$ is
responsible for calculating its own marginal cost $\frac{\partial
D}{\partial r_i(w)}$ according to
\eqref{eq:NodeMarginalRoutingCost2}, and then providing
$\frac{\partial D}{\partial r_i(w)}$ to {its neighbors from which it
receives traffic of $w$}. In~\cite{paper:Gal77}, the rules for
propagating the marginal routing cost information are specified.

\vspace{3mm} {\bf Loop-Free Routing and Blocked Node Sets}: The
existence of loops in a routing pattern gives rise to redundant
circulation of data flows, hence wasting network resources.  The
device of blocked node sets ${\cal B}_i(w)$ was invented in
\cite{paper:Gal77,paper:BGG84} to suppress the formation of loops in
each iteration of the distributed routing algorithm. Intuitively,
the blocking mechanism works as follows. A node does not forward
flow to a neighbor with higher marginal cost {or to a neighbor that
routes positive flow to some other node with higher marginal cost}.
Such a scheme guarantees that each session's traffic flows through
nodes in decreasing order of marginal costs, thus precluding the
existence of loops. For more details, please refer
to~\cite{paper:Gal77,paper:BGG84}.


\vspace{3mm}

{\bf Scaling Matrices and Stepsizes:}
Generally speaking, there is a tradeoff between the complexity of
algorithm iterations and the speed of convergence to the optimal
point.  A simple structure for the scaling matrix can greatly reduce
the complexity of each iteration. In particular, if
\be\label{eq:BasicRTScalingMatrix} M_i^k(w) =
\frac{t_i^k(w)}{\alpha_i^k(w)} \cdot {\rm diag}\{1, \cdots,1, 0, 1,
\cdots, 1\},\ee where the only zero entry on the diagonal is at the
$j$th place such that $j \in \arg\min_{l}\delta\phi_{il}^k(w)$,
then~\eqref{eq:GeneralRTUpdate} becomes equivalent to the routing
algorithm by Gallager~\cite{paper:Gal77}. That is
\begin{equation}\label{eq:BasicRT}
\boldsymbol\phi_i^{k+1}(w) = \boldsymbol\phi_i^k(w) + \Delta\boldsymbol\phi_i(w),
\end{equation}
where the increment $\Delta\boldsymbol\phi_i(w) = (\Delta\phi_{ij}(w))_{j \in {\cal
O}(i)}$ is given by
\begin{equation}\label{eq:BasicRTUpdate}
\begin{array}{ll}
\vspace{3mm} \Delta\phi_{ij}(w) = 0, \; &\forall j \in {\cal B}_i^k(w),
\\
\vspace{1mm} a_{ij} \triangleq \delta\phi_{ij}^k(w) - \displaystyle\min_{l \in {\cal
O}(i) \backslash {\cal B}_i^k(w)} \delta\phi_{il}^k(w), &\forall j \in {\cal O}(i)
\backslash {\cal B}_i^k(w),
\\
\vspace{1mm} \Delta\phi_{ij}(w) =  -\min\left\{\phi_{ij}^k(w),
\displaystyle\frac{\alpha_i^k(w) a_{ij}}{t_i^k(w)}\right\}, \; &\forall j:a_{ij} > 0,
\\
\vspace{1mm} \Delta\phi_{ij}(w) =  -\displaystyle\sum_{l \ne j}{\Delta\phi_{il}(w)},
&\textrm{for one}~j:a_{ij} = 0.
\\
\end{array}
\end{equation}
We will refer to \eqref{eq:BasicRT}-\eqref{eq:BasicRTUpdate} as the
basic routing algorithm or BRT.  The BRT simplifies the quadratic
optimization in~\eqref{eq:GeneralRTUpdate} to a scalar form and
reduces the scaling matrix selection to a choice of the stepsize
$\alpha_i^k(w)$.  The simplicity of a BRT iteration, however, comes
at the expense of the convergence rate.   In particular, excessively
small stepsizes can lead to slow convergence.  This is the case for
the routing algorithm of Gallager~\cite{paper:Gal77}, for which the
stepsizes are proportional to $|{\cal N}|^{-6}$).

In order to improve the convergence rate, the scaling matrix
$M_i^k(w)$ needs to approximate the Hessian more closely.  This is
the approach adopted in~\cite{paper:BGG84}, where second-derivative
algorithms are developed.  The scaling matrix is obtained by
dropping all off-diagonal terms of the Hessian matrix, and
approximating the diagonal terms via a second-derivative information
exchange process~\cite{paper:BGG84}.  Here, each iteration entails a
more complex quadratic program.  The Hessian approximation scheme
in~\cite{paper:BGG84} is quite involved.  Moreover, the algorithm
works well only near the optimum.  When starting from a point far
from the optimum, convergence cannot be guaranteed.  This is due to
the fact that the scaling matrices generally are not upper bounds on
the Hessians, and the Hessians being estimated are evaluated at the
current routing configuration rather than at intermediate points
between the current and next routing configurations.

\vspace{3mm}

\subsubsection{A New Scaled Gradient Projection Routing Algorithm}

In this section, we present a scaled gradient projection routing
algorithm for wireless networks based on a new scaling matrix
selection scheme.  In this new scheme, the scaling matrix is chosen
to be an upper bound on the Hessian matrix evaluated at any
intermediate point between the current and next routing
configuration.  The new scheme has several advantages over the
approach of~\cite{paper:Gal77} and~\cite{paper:BGG84}.  First, our
technique can generate stepsizes for the BRT algorithm
of~\cite{paper:Gal77} which are larger than those
in~\cite{paper:Gal77}, leading to an improved convergence rate.
Second, in contrast to the approximation scheme used
in~\cite{paper:BGG84}, our method requires less control overhead for
distributed computation. More importantly, since our scheme finds an
upper bound on the Hessian matrices evaluated at any intermediate
configuration, it {\em guarantees convergence} of the GRT from any
initial point.  Finally, whereas the algorithms
in~\cite{paper:Gal77} and~\cite{paper:BGG84} assume that all nodes
in the network iterate at the same time, our algorithms allows nodes
to update one at a time. This latter mode of operation may be more
appropriate in wireless networks without a central controller, where
individual nodes can update their routing variables only in an
autonomous and asynchronous manner.


To describe our new algorithm, let ${\cal AN}_i^k(w) \triangleq
{\cal O}(i) \backslash {\cal B}_i^k(w)$ and let $h_i^k(w)$ denote
the maximum number of hops on a path from $i$ to $D(w)$. Given that
the initial network cost is upper bounded by $D^0 < \infty$, node
$i$ finds the quantities
\[
A_{ij}^k(D^0) \triangleq \max_{F_{ij}:D_{ij}(C_{ij}^k, F_{ij}) \le D^0} \frac{\partial^2
D_{ij}}{\partial F_{ij}^2},\]
\[
A^k(D^0) \triangleq \max_{(m,n)\in {\cal E}} A^k_{mn}(D^0).
\]
A diagonal upper bound on the Hessian matrix with respect to the
routing variables can be found as in the following crucial lemma.
Its proof is contained in Appendix~\ref{app:PhiHessian}.
\vspace{0.1in}\begin{lemma}\label{lma:PhiHessianUppBnd} If the
initial network cost is less than or equal to $D^0 < \infty$, then
at every iteration $k$ of the general routing algorithm
of~\eqref{eq:GeneralRTUpdate} and for all $\lambda \in [0,1]$,
$H_{\boldsymbol\phi_i(w)}^{k,\lambda} \triangleq \nabla^2
D(\boldsymbol\phi_i (w))|_{\lambda\boldsymbol\phi_i^k
(w)+(1-\lambda)\boldsymbol\phi_i^{k+1} (w)}$ is upper bounded by the
diagonal matrix
\[ \bar M_i^k(w) = {t_i^k(w)}^2{\rm diag}\left\{
 \left(A^k_{ij}(D^0) + |{\cal AN}^k_i(w)| h_j^k(w) A^k(D_0)\right)_{j \in {\cal AN}_i^k(w)}\right\}\]
 in the sense that for all $\boldsymbol v_i \in {\cal V}_i^k(w)
 = \left\{ \boldsymbol v_i : \sum_{j \in {\cal AN}_i^k(w)}{v_{ij}} = 0\right\}$, $\boldsymbol v_i' \cdot H_{\boldsymbol\phi_i (w)}^{k,\lambda} \cdot \boldsymbol v_i
\leq \boldsymbol v_i' \cdot \bar M_i^k(w)  \cdot \boldsymbol v_i$.
\end{lemma}\vspace{0.1in}

\textit{Proof:} See Appendix~\ref{app:PhiHessian}. \vspace{0.1in}

Note that the evaluation of $\bar M_i^k(w)$ requires a simple
protocol in which at each iteration~$k$, each node $j$ provides its
immediate upstream neighbors with $h_j^k(w)$, which is derived from
those counts reported by $j$'s next-hop neighbors. The computation
can be carried out in a simple distributed Bellman-Ford form:
\[
h_j^k(w) = \max_{l \in {\cal O}(j)} h_l^k(w) + 1,
\]
where $h_{D(w)}^k(w) \equiv 0$.

{We will show that if we choose $2 t_i^k(w) M_i^k(w)$ to closely
upper bound $H_{\boldsymbol\phi_i(w)}^{k,\lambda}$ via
Lemma~\ref{lma:PhiHessianUppBnd}, the resulting routing algorithms
will have fast and guaranteed convergence to the optimal
configuration. For the
BRT~\eqref{eq:BasicRT}-\eqref{eq:BasicRTUpdate}, this amounts to
choosing the stepsize $\alpha_i^k(w)$ as
\begin{equation}\label{eq:BasicRTStepSizeRange}
\alpha_i^k(w) = 2 \left[|{\cal AN}_i^k(w)| \max_{j \in {\cal AN}_i^k(w)}\left\{A^k_{ij}(D_0) +
|{\cal AN}_i^k(w)| h_j^k(w) A^k(D_0)\right\}\right]^{-1}.
\end{equation}

{For the GRT, this amounts to choosing the scaling matrix $M_i^k(w)$
as \be\label{eq:RTscaling} M_i^k(w) = \frac{\bar M_i^k(w)}{2
t_i^k(w)} = \frac{t_i^k(w)}{2} {\rm diag}\left\{ \left(A_{ij}^k(D^0)
+ |{\cal AN}_i^k(w)| h_j^k(w) A^k(D^0) \right)_{j \in {\cal
AN}_i^k(w)} \right\}. \ee As we will show later in
Theorem~\ref{thm:AlgorithmConvergence}, with $\alpha_i^k(w)$ and
$M_i^k(w)$ specified above, each iteration of BRT or GRT strictly
reduces the network cost unless conditions
\eqref{eq:RoutingOptCond1}-\eqref{eq:RoutingOptCond2} are satisfied
by $\delta\bs\phi_i^k(w)$.}

\subsection{Power Allocation Algorithm (PA)}

Let $PA(\boldsymbol\eta_i)$ denote the algorithm applied by node $i$ to vary its transmission power allocation
variables. At the $k$th iteration, $PA$ updates the current local power allocation $\boldsymbol\eta_i^k =
(\eta_{ij}^k)_{j \in {\cal O}(i)}$ by $\boldsymbol\eta_i^{k+1} = PA(\boldsymbol\eta_i^k)$ where
$\boldsymbol\eta_i^{k+1}$ is the solution to
\begin{equation}\label{eq:GeneralPAUpdate}
\begin{array}{ll}
\vspace{1mm} \textrm{minimize} \; &\displaystyle{\delta{\boldsymbol\eta_i^k}' \cdot (\boldsymbol\eta_i -
\boldsymbol\eta_i^k) + \frac{1}{2}(\boldsymbol\eta_i - \boldsymbol\eta_i^k)' \cdot Q_i^k \cdot (\boldsymbol\eta_i -
\boldsymbol\eta_i^k)}
\\
\textrm{subject to} &\boldsymbol\eta_i \ge \boldsymbol 0, ~ \sum\limits_{j \in {\cal O}(i)}{\eta_{ij}} = 1.
\end{array}
\end{equation}
We refer to~\eqref{eq:GeneralPAUpdate} as the general power allocation algorithm or GPA. Here,
$\delta\boldsymbol\eta_i^k \triangleq (\delta\eta_{ij}^k)_{j \in {\cal O}(i)}$, and $Q_i^k$ is the
scaling matrix, which we will specify in a moment.
\subsubsection{Local Message Exchange}

Note that marginal power allocation costs $\delta\eta_{ij}$ involve \emph{only locally obtainable
measures} (cf. \eqref{eq:MarginalPowerAlloc}). Thus, the power allocation algorithm needs only a
simple local message exchange before an iteration of $PA$.

In particular, let each neighbor $j$ of node $i$ measure the value of $SINR_{ij}$ and feed it back
to $i$. Then $i$ can readily compute all $\delta\eta_{ij}$'s according to
\[
\delta\eta_{ij} = \frac{\partial D_{ij}}{\partial C_{ij}} \frac{C_{ij}' SINR_{ij}}{P_{ij}}(1 +
SINR_{ij}),
\]
which follows from a modification of \eqref{eq:MarginalPowerAlloc}.

\subsubsection{Scaling Matrix}

As in the BRT of Gallager, we can adopt a simple structure for
$Q_i^k$ to facilitate iterations at each node.  Specifically, let
$Q_i^k = Q/\beta_i^k$ where $\beta_i^k$ is a positive scalar and $Q
= P_i^k \rm{diag}\{1,\cdots,1,0,1,\cdots,1\}$ with the only zero
entry at the $j$th place such that $j \in
\arg\min_{l}\delta\eta_{il}^k$. Thus, the
GPA~\eqref{eq:GeneralPAUpdate} is reduced to the following basic
power allocation algorithm (BPA):
\begin{equation}\label{eq:BasicPA}
\boldsymbol\eta_i^{k+1} = \boldsymbol\eta_i^k + \Delta\boldsymbol\eta_i,
\end{equation}
where the increment $\Delta\boldsymbol\eta_i = (\Delta\eta_{ij})_{j \in {\cal O}(i)}$ is computed
as
\begin{equation}\label{eq:BasicPAUpdate}
\begin{array}{ll}
\vspace{1mm} \displaystyle{b_{ij} \triangleq \delta\eta_{ij}^k - \min_{l \in {\cal O}(i)}
\delta\eta_{il}^k},
\\
\vspace{1mm} \Delta\eta_{ij} = -\min\{\eta_{ij}^k,\beta_i^k b_{ij} / P_i \}, \; &\forall j:b_{ij} >
0,
\\
\vspace{1mm} \Delta\eta_{ij} = -\displaystyle\sum_{l:b_{il} > 0}{\Delta\eta_{il}}, &\textrm{for
one}~j:b_{ij} = 0.
\end{array}
\end{equation}

We now specify the appropriate stepsize $\beta_i^k$ for BPA and
appropriate scaling matrix $Q_i^k$ for the GPA. Assume that the sum
of the local link costs at node $i$ before the $k$th iteration is
$\sum_{j \in {\cal O}(i)} D_{ij}^k = D_i^k$. Since the powers used
by the other nodes do not change over the iteration, $C_{ij}$
depends only on $\eta_{ij}$ as
\[
C_{ij} = C(x_{ij}) = C\left(\frac{ G_{ij} P_i \eta_{ij}}{G_{ij} P_{i} (1 - \eta_{ij}) + \sum_{m \ne i} G_{mj} P_{m} +
N_j}\right) \triangleq C_{ij}(\eta_{ij}).
\]
It can be shown that there exists a lower bound $\underline\eta_{ij}$ on the updated value of $\eta_{ij}$ such that
$\underline C_{ij} = C_{ij}(\underline\eta_{ij})$ and $D_{ij}(\underline C_{ij}, F_{ij}^k) = D_i^k$. Accordingly, the
possible range of $x_{ij}$ is \[ x_{ij}^{min} \triangleq \frac{ G_{ij} P_i \underline\eta_{ij}} {G_{ij} P_{i} (1 -
\underline\eta_{ij}) + \sum_{m \ne i} G_{mj} P_{m} + N_j} \le x_{ij} \le \frac{G_{ij} P_i} {\sum_{m \ne i} G_{mj} P_{m}
+ N_j} \triangleq x_{ij}^{max}. \] Define an auxiliary term $\beta_{ij}$ as \begin{small}\be\label{eq:beta} \beta_{ij}
= \frac{1}{\underline\eta_{ij}^2}  \left[ B_{ij}(D_i^k)  \max_{x_{ij}^{min} \le x \le x_{ij}^{max}} \{ C'(x)^2 x^2 (1 +
x)^2 \}  + \left.\frac{\partial D_{ij}}{\partial C_{ij}} \right|_{D_{ij}(C_{ij}, F_{ij}^k) = D_i^k} \cdot
\min_{x_{ij}^{min} \le x \le x_{ij}^{max}} \{ C''(x) x^2 (1 + x)^2 \} \right] \ee
\end{small}where $B_{ij}(D_i^k) = \max_{D_{ij}(C_{ij}, F_{ij}^k) \le D_i^k}
\frac{\partial^2 D_{ij}}{\partial C_{ij}^2}$. We have the following
important lemma, whose proof is deferred to
Appendix~\ref{app:EtaHessian}.

\vspace{0.1in}\begin{lemma}\label{lma:EtaHessianUppBnd} Denote the
local cost at node $i$ at the beginning of iteration $k$ of the
power allocation algorithm by $D_i^k \triangleq \sum_{j \in {\cal
O}(i)} D_{ij}^k$, then for all $\lambda \in [0,1]$, the Hessian
matrix $H_{\boldsymbol\eta_i}^{k,\lambda} \triangleq \nabla^2
D(\boldsymbol\eta_i)|_{\lambda\boldsymbol\eta_i^k+(1-\lambda)\boldsymbol\eta_i^{k+1}}$
is upper bounded by the diagonal matrix
\[
\bar Q_i^k = {\rm diag}\{(\beta_{ij})_{j \in {\cal O}(i)}\}
\]
with $\beta_{ij}$ given by \eqref{eq:beta}, in the sense that for all $\boldsymbol v_i \in {\cal V}_{\bs\eta_i}
\triangleq \left\{ \boldsymbol y_i : \sum_{j \in {\cal O}(i)}{y_{ij}} = 0\right\}$, $\boldsymbol v_i' \cdot
H_{\boldsymbol\eta_i}^{k,\lambda} \cdot \boldsymbol v_i \leq \boldsymbol v_i' \cdot \bar Q_i^k \cdot \boldsymbol v_i$.
\end{lemma}\vspace{0.1in}
Using Lemma~\ref{lma:EtaHessianUppBnd}, we can choose the stepsize
$\beta_i^k$ in the BPA algorithm to be
\be\label{eq:BasicPAStepsizeRange} \beta_i^k = 2 (P_i^k)^2
\left[|{\cal O}(i)|\max_{j \in {\cal O}(i)} \beta_{ij} \right]^{-1},
\ee and the scaling matrix $Q_i^k$ for the GPA algorithm to be
\be\label{eq:PAscaling} Q_i^k =\frac{\bar Q_i^k}{2 P_i^k} =
\frac{1}{2 P_i^k} {\rm diag} \left\{ (\beta_{ij})_{j \in {\cal
O}(i)} \right\}. \ee {It can be shown using the arguments of
Theorem~\ref{thm:AlgorithmConvergence} below that the BPA and GPA
algorithms with the $\beta_i^k$ and $Q_i^k$ specified above strictly
reduce the network cost at every iteration unless
\eqref{eq:PowerOptCond1} is satisfied by $\delta\bs\eta_i^k$.}

\subsection{Power Control Algorithm (PC)}

At the $k$th iteration of the power control algorithm $PC$, the power control variables $\boldsymbol\gamma^k =
(\gamma_i^k)_{i \in {\cal N}}$ are updated by
\begin{equation}\label{eq:GeneralPC}
\boldsymbol\gamma^{k+1} = PC(\boldsymbol\gamma^k),
\end{equation}
where $\boldsymbol\gamma^{k+1}$ is the solution to
\begin{equation}\label{eq:GeneralPCUpdate}
\begin{array}{ll}
\vspace{1mm} \textrm{minimize} \; &\displaystyle{\delta{\boldsymbol\gamma^k}' \cdot (\boldsymbol\gamma -
\boldsymbol\gamma^k) + \frac{1}{2}(\boldsymbol\gamma - \boldsymbol\gamma^k)' \cdot V^k \cdot (\boldsymbol\gamma -
\boldsymbol\gamma^k)}
\\
\textrm{subject to} &\boldsymbol\gamma \le \mathbf 1.
\end{array}
\end{equation}
Here matrix $V^k$ is symmetric, positive definite on $\mathbb R^{|{\cal N}|}$. Note that in general
\eqref{eq:GeneralPCUpdate} represents a coordinated network-wide algorithm. It can be decomposed into distributed
computations if and only if $V^k$ is diagonal. In this case, denote $V^k = {\rm diag} \{ (v_i)_{i \in {\cal N}} \}$,
\eqref{eq:GeneralPCUpdate} is then transformed to $|{\cal N}|$ parallel local sub-programs, each having the form
\begin{equation}\label{eq:BasicPC}
\gamma_i^{k+1} = PC(\gamma_i^k) = \min \left\{ 1, \gamma_i^k - \frac{\delta\gamma_i^k}{v_i} \right\}.
\end{equation}

\subsubsection{Power Control Message Exchange}\label{subsec:PCMsgEx}

Unlike the power allocation algorithm, $\delta\gamma_i$ depends on external information
from nodes $m \ne i$ (cf.~\eqref{eq:MarginalPowerCtrl}). Thus, its calculation must be
preceded by a message exchange phase. Before introducing the message exchange protocol,
we re-order the summations on the RHS of \eqref{eq:MarginalPowerCtrl} as
\be\label{eq:DeltaGamma} \frac{\delta\gamma_i}{P_i}  = \sum_{n \in {\cal N}}\left[ G_{in}
\sum_{m \in {\cal I}(n)} -{\frac{\partial D_{mn}}{\partial C_{mn}} \frac{C_{mn}'
SINR_{mn}}{IN_{mn}}}\right] + \sum_{n \in {\cal O}(i)} \delta\eta_{in} \cdot \eta_{in}.
\ee With reference to the expression above, we propose the following protocol for
computing the values of $\delta\gamma_i$ for all $i \in {\cal N}$.

\emph{Power Control Message Exchange Protocol:} Let each node $n$ assemble the measures \[{- \frac{\partial
D_{mn}}{\partial C_{mn}} \frac{C_{mn}' SINR_{mn}}{IN_{mn}}} = {-\frac{\partial D_{mn}}{\partial C_{mn}} \frac{C_{mn}'
SINR_{mn}^2}{G_{mn} P_{mn}}}\] on all its incoming links $(m,n)$, and sum them up to form the
\be\label{eq:MSG}\text{Power Control Message:} \quad MSG(n) = \sum_{m \in {\cal I}(n)}{-\frac{\partial D_{mn}}{\partial
C_{mn}} \frac{C_{mn}' SINR_{mn}^2}{G_{mn} P_{mn}}}.\ee It then broadcasts $MSG(n)$ to the whole network via a flooding
protocol. This control message generating process is illustrated by Figure \ref{fig:PC}.
\begin{figure}
\begin{center}
\includegraphics[width = 10cm]{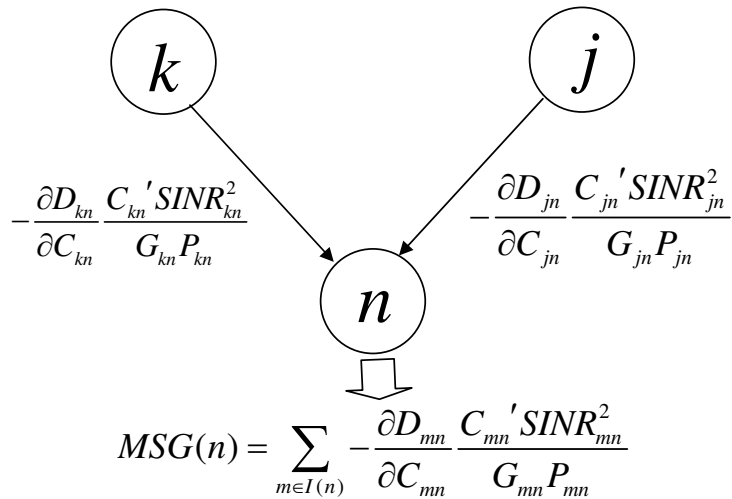}
\caption{Power Control Message Generation}\label{fig:PC}
\end{center}
\end{figure}
Upon obtaining $MSG(n)$, node $i$ processes it according to the following rule. If $n$ is
a next-hop neighbor of $i$, node $i$ multiplies $MSG(n)$ with path gain 
$G_{in}$ and adds the product to the value of local measure
$\delta\eta_{in} \cdot \eta_{in}$; otherwise, node $i$ multiplies
$MSG(n)$ with $G_{in}$. Finally, node $i$ adds up all the processed
messages, and this sum multiplied by $P_i$  equals $\delta\gamma_i$.
Note that this protocol requires {\em only one message from each
node in the network}. Moreover in practice, a node $i$ can
effectively ignore the messages generated by distant nodes. To see
this, note that messages from distant nodes contribute very little
to $\delta\gamma_i$ due to the negligible multiplicative factor
$G_{in}$ on $MSG(n)$ when $i$ and $n$ are far apart
(cf.~\eqref{eq:DeltaGamma}).  This observation is borne out by the
results of numerical simulations presented in
Section~\ref{sec:Simulation}, where it is shown that the power
control algorithm converges reasonably well even when every node
exchanges power control messages only with its close neighbors.

\subsubsection{Alternative Implementation}

Note that it is not mandatory to have all the nodes $i \in {\cal N}$
perform an update at each instance of the $PC(\cdot)$ algorithm. One
may consider the case where only a subset of nodes ${\cal N}^k$
iterate $PC(\cdot)$, i.e. $\gamma_i^{k+1} = PC(\gamma_i^k)$ for all
$i \in {\cal N}^k$, and $\gamma_i^{k+1} = \gamma_i^k $ for all $i
\notin {\cal N}^k$. As long as no node is left out of the updating
set ${\cal N}^k$ indefinitely when the conditions
\eqref{eq:PowerOptCond2}-\eqref{eq:PowerOptCond3} are not satisfied
by $\gamma_i$, the convergence result proved in the following
subsection applies. However, in order to minimize control messaging
overhead, it may be preferable to have each round of global power
control message ($MSG(n)$) exchange induce one iteration of power
control algorithm at every node (as opposed to iterations at only a
subset of nodes). Our subsequent analysis of algorithm convergence
and scaling matrix selection will be based on this latter mode of
implementation.

\subsubsection{Scaling Matrix}

As for previous algorithms, we select the scaling matrix $V^k$ to be a diagonal upper
bound on the Hessian matrix. Specifically, given that the initial network cost is less
than or equal to $D^0$, the following terms can be evaluated:
\[
\bar B(D^0) = \max_{(m,n) \in {\cal E}} \max_{D_{mn} \le D^0} \frac{\partial^2 D_{mn}}{\partial C_{mn}^2},
\]
\[
\underline B(D^0) = \min_{(m,n) \in {\cal E}} \min_{D_{mn} \le D^0} \frac{\partial D_{mn}}{\partial C_{mn}}.
\]
Moreover, due to the individual power constraints~\eqref{eq:TotalPowerConstraint}, there exists a finite upper bound
$\bar x$ on the achievable $SINR$ on all links. Define $\kappa \triangleq \max_{0 \le x \le \bar x} {C'(x)}^2 \cdot
x^2$, and $\varphi \triangleq \min_{0 \le x \le \bar x} C''(x) \cdot x^2$.

\vspace{0.1in}\begin{lemma}\label{lma:GammaHessianUppBnd}  Assume
the initial network cost is less than or equal to $D^0 < \infty$. At
each iteration of the power control algorithm
\eqref{eq:GeneralPCUpdate}, $H_{\boldsymbol\gamma}^{k,\lambda}
\triangleq \nabla^2
D(\boldsymbol\gamma)|_{\lambda\boldsymbol\gamma^k+(1-\lambda)\boldsymbol\gamma^{k+1}}$
is upper bounded by the diagonal matrix
\[ \bar V = |{\cal N}| |{\cal E}| \left[ \bar B(D^0)
\kappa + \underline B(D^0) \varphi \right] {\rm diag} \left\{ \left( \bar S_i^2 \right)_{i \in {\cal N}} \right\} \]
for all $\lambda \in [0,1]$, in the sense that for all $\boldsymbol v \in \mathbb R^{|{\cal N}|}$,
\[
\boldsymbol v' \cdot H_{\boldsymbol\gamma}^{k,\lambda}
 \cdot
\boldsymbol v \leq \boldsymbol v' \cdot \bar V \cdot \boldsymbol v.
\]
\end{lemma}\vspace{0.1in}

The proof of the lemma is contained in
Appendix~\ref{app:GammaHessian}. Using
Lemma~\ref{lma:GammaHessianUppBnd} and the arguments of
Theorem~\ref{thm:AlgorithmConvergence} below, we can show that with
the scaling matrices chosen as \be\label{eq:PCscaling} V =
\frac{1}{2} |{\cal N}| |{\cal E}| \left[ \bar B(D^0) \kappa +
\underline B(D^0) \varphi \right] {\rm diag}\{(\bar S_i)_{i \in
{\cal N}}\}, \ee the power control algorithm strictly reduces the
network cost at every iteration unless
\eqref{eq:PowerOptCond2}-\eqref{eq:PowerOptCond3} are satisfied by
$\delta\bs\gamma^k$.}

Notice that $V$ is independent of the iteration index $k$, and can be determined at the first
iteration. {Also notice that applying the scaling matrix $V$ in \eqref{eq:PCscaling} is equivalent
to letting each node set \[ v_i = \frac{\bar S_i}{2} |{\cal N}| |{\cal E}| \left[ \bar B(D^0)
\kappa + \underline B(D^0) \varphi \right]
\] in the node-based iteration~\eqref{eq:BasicPC}.}

\subsection{Convergence of Algorithms}

\label{sec:ConvergenceProof}

We now prove the central convergence result for the class of
scaled gradient projection algorithms discussed above.
\vspace{0.1in}

\begin{theorem}\label{thm:AlgorithmConvergence}
Assume an initial loop-free routing configuration $(\boldsymbol\phi_i^0(w))$ and initial valid transmission power
configuration $(\boldsymbol\eta_i^0)$ and $\boldsymbol\gamma^0$ such that the initial network cost is upper bounded by
 $D^0 < \infty$. Then the sequences generated by the BRT, BPA algorithms with stepsizes given by~\eqref{eq:BasicRTStepSizeRange} and \eqref{eq:BasicPAStepsizeRange} or by the GRT, GPA and PC algorithms with scaling matrices given by~\eqref{eq:RTscaling}, \eqref{eq:PAscaling} and \eqref{eq:PCscaling} converge, i.e., $\{\boldsymbol\phi_i^k(w)\} \to
\{\boldsymbol\phi_i^*(w)\}$, $\{\boldsymbol\eta_i^k\} \to \{\boldsymbol\eta_i^*\}$, and
$\boldsymbol\gamma^k \to \boldsymbol\gamma^*$ as $k \to \infty$. Furthermore, the limits
$\{\boldsymbol\phi_i^*(w)\}$, $\{\boldsymbol\eta_i^*\}$ and $\boldsymbol\gamma^*$ satisfy the
optimality conditions~\eqref{eq:RoutingOptCond1}-\eqref{eq:PowerOptCond3}.
\end{theorem}\vspace{0.1in}

\textit{Proof:} We first show that with the stepsizes and scaling
matrices specified earlier, every iteration of each algorithm
strictly reduces the network cost unless the corresponding
equilibrium conditions
in~\eqref{eq:RoutingOptCond1}-\eqref{eq:PowerOptCond3} of the
adjusted variables are satisfied. We present a detailed proof {for
the stepsizes and scaling matrices in the basic and general routing
algorithms} $RT(\boldsymbol\phi_i^k(w))$. The analysis for the other
algorithms is almost verbatim. For notational convenience, the
session index $w$ is suppressed.

Consider the $k$th iteration of $RT(\cdot)$. If $t_i^k = 0$, the algorithm has no effect on the network cost whatever
the update is. 
We thus focus on the case of $t_i^k > 0$. Since $M_i^k$ is positive definite, the objective
function of \eqref{eq:GeneralRTUpdate} is convex in $\boldsymbol\phi_i$. Moreover, since the
feasible set ${\cal F}_i^k$ is convex, the solution $\boldsymbol\phi_i^{k + 1}$ satisfies
\cite{book:Ber99}
\begin{equation}\label{eq:ConvexOptimality}
\left[\delta\boldsymbol\phi_i^k + {M_i^k}(\boldsymbol\phi_i^{k + 1} - \boldsymbol\phi_i^k)\right]' \cdot
(\boldsymbol\phi_i^{k + 1} - \boldsymbol\phi_i) \le 0, \; \forall \boldsymbol\phi_i \in {\cal F}_i^k.
\end{equation}
Setting $\boldsymbol\phi_i = \boldsymbol\phi_i^k$, we obtain
\begin{equation}\label{eq:ConvexOptimalityCont}
{\delta\boldsymbol\phi_i^k}' \cdot (\boldsymbol\phi_i^{k + 1} - \boldsymbol\phi_i^k) \le -(\boldsymbol\phi_i^{k + 1} -
\boldsymbol\phi_i^k)' \cdot {M_i^k} \cdot (\boldsymbol\phi_i^{k + 1} - \boldsymbol\phi_i^k).
\end{equation}
By Taylor's Expansion, the network cost difference after the current iteration is
\begin{equation}\label{eq:CostDiffAfterIteration}
\begin{array}{ll}
\vspace{1mm} D(\boldsymbol\phi_i^{k + 1}) - D(\boldsymbol\phi_i^k) &= (t_i^k \cdot \delta\boldsymbol\phi_i^k)' \cdot
(\boldsymbol\phi_i^{k + 1} - \boldsymbol\phi_i^k) + \displaystyle\frac{1}{2} (\boldsymbol\phi_i^{k + 1} -
\boldsymbol\phi_i^k)' \cdot H_{\boldsymbol\phi_i}^{k,\lambda} \cdot (\boldsymbol\phi_i^{k + 1} - \boldsymbol\phi_i^k)
\\
&\le (\boldsymbol\phi_i^{k + 1} - \boldsymbol\phi_i^k)' \cdot \left(\displaystyle{- {t_i^k M_i^k}  +
\frac{H_{\boldsymbol\phi_i}^{k,\lambda}}{2}}\right) \cdot (\boldsymbol\phi_i^{k + 1} - \boldsymbol\phi_i^k),
\end{array}
\end{equation}
where $H_{\boldsymbol\phi_i}^{k,\lambda}$ is the Hessian matrix of $D$ with respect to
components of $\boldsymbol\phi_i$, evaluated at $\lambda \boldsymbol\phi_i^k +
(1-\lambda) \boldsymbol\phi_i^{k+1}$ for some $\lambda \in [0,1]$. By
Lemma~\ref{lma:PhiHessianUppBnd}, both the $M_i^k$ given
by~\eqref{eq:BasicRTScalingMatrix} with $\alpha_i^k$ given
by~\eqref{eq:BasicRTStepSizeRange} and the $M_i^k$ given by~\eqref{eq:RTscaling} upper
bound $H_{\boldsymbol\phi_i}^{k,\lambda} / (2 t_i^k)$ in the sense that $- {t_i^k M_i^k}
+ {H_{\boldsymbol\phi_i}^{k,\lambda}}/{2}$ is negative definite. Thus, with one iteration
$D(\boldsymbol\phi_i^{k + 1}) - D(\boldsymbol\phi_i^k) \le 0$, where the inequality is
strict unless $\boldsymbol\phi_i^{k+1} = \boldsymbol\phi_i^k$, which happens only when
conditions \eqref{eq:RoutingOptCond1}-\eqref{eq:RoutingOptCond2} hold at
$\boldsymbol\phi_i^k$. In conclusion, an iteration of BRT with $\alpha_i^k$
in~\eqref{eq:BasicRTStepSizeRange} or an iteration of GRT with $M_i^k$
in~\eqref{eq:RTscaling} strictly reduces the network cost until the equilibrium
conditions for $\bs\phi_i$ are satisfied.

Similarly, by Lemmas~\ref{lma:EtaHessianUppBnd} and \ref{lma:GammaHessianUppBnd}, we can show that
network cost is strictly reduced by the iterations of the BPA, GPA and PC algorithms with stepsizes
or scaling matrices given by~\eqref{eq:BasicPAStepsizeRange}, \eqref{eq:PAscaling} and
\eqref{eq:PCscaling}, unless \eqref{eq:PowerOptCond1}-\eqref{eq:PowerOptCond3} are satisfied by the
current $\bs\eta_i^k$ and $\bs\gamma^k$.

To summarize, with the specific choices of stepsizes and scaling
matrices derived earlier, any iteration of any of the algorithms
BRT, GRT, BPA, GPA and PC strictly reduces the total network cost
with all other variables fixed, unless the equilibrium conditions
for the adjusted optimization variables
(\eqref{eq:RoutingOptCond1}-\eqref{eq:RoutingOptCond2} for
$\bs\phi_i(w)$, \eqref{eq:PowerOptCond1} for $\bs\eta_i$,
\eqref{eq:PowerOptCond2}-\eqref{eq:PowerOptCond3} for $\gamma_i$)
are satisfied. Recall that the feasible sets of $\bs\phi_i(w)$,
$\bs\eta_i$ and $\bs\gamma$ are given
by~\eqref{eq:RoutingVariableConstraint} and
\eqref{eq:PowerVariableConstraint}.  The sequences
$\{\boldsymbol\phi_i^k(w)\}_{k=0}^{\infty}$ and
$\{\boldsymbol\eta_i^k\}_{k=0}^{\infty}$ clearly take values in
compact sets. Although $\bs\gamma^k$ is explicitly only upper
bounded by $\bs 1$, the fact that the network cost is always upper
bounded by $D^0$ implies an implicit lower bound on
$\bs\gamma$.\footnote{For each component $\gamma_i$ of $\bs\gamma$,
a lower bound can be derived as $\underline\gamma_i = \max_{j \in
{\cal O}(i)}\underline\gamma_{ij}$ where $D_{ij}(C((G_{ij}(\bar
P_i)^{\underline\gamma_{ij}})/N_j),0) = D^0$.  That is,
$\underline\gamma_{ij}$ is the power control level that yields a
cost of $D^0$ on link $(i,j)$ assuming the total power of $i$ is
allocated exclusively to $(i,j)$ and all other links are
non-interfering.} Thus, for any finite initial network cost $D^0$,
$\{\boldsymbol\gamma^k\}_{k=0}^{\infty}$ also takes values in a
compact set.  It follows that
$\{\boldsymbol\phi_i^k(w)\}_{k=0}^{\infty}$,
$\{\boldsymbol\eta_i^k\}_{k=0}^{\infty}$, and
$\{\boldsymbol\gamma^k\}_{k=0}^{\infty}$ must each have a convergent
subsequence. Since the sequence of network costs generated by
iterations of all the algorithms is non-increasing and bounded
below, it must have a limit $D^*$. Therefore, the network cost at
the limit points $\boldsymbol\phi_i^*(w)$, $\boldsymbol\eta_i^*$ and
$\boldsymbol\gamma^*$ of the convergent subsequences must coincide
with $D^*$. Because $D^*$ cannot be further (strictly) reduced by
the algorithm iterations, $\boldsymbol\phi_i^*(w)$,
$\boldsymbol\eta_i^*$ and $\boldsymbol\gamma^*$ must satisfy
conditions \eqref{eq:RoutingOptCond1}-\eqref{eq:PowerOptCond3}. \qed


From the proof we can see that the global convergence does not
require any particular order in running the three algorithms at
different nodes. For convergence to the joint optimum, every node
$i$ only needs to iterate its own algorithms until its routing,
power allocation, and power control variables satisfy
\eqref{eq:RoutingOptCond1}-\eqref{eq:PowerOptCond3}.\footnote{In
practice, nodes may keep updating their optimization variables with
the corresponding algorithms until further reduction in network cost
by any one of the algorithms is negligible.}

It is important to note that the structure of the routing, power
allocation, and power control algorithms make them particularly
desirable for distributed implementation without knowledge of global
network topology or traffic patterns.  The algorithms are
fundamentally driven by the relevant marginal cost messages.  These
marginal cost messages contain all the information regarding the
whole network which is relevant to each iteration of any algorithm
at any given node.  Thus, it is not necessary for the network to
perform localization or traffic matrix estimation in order carry out
optimal routing.  The fact that the algorithms are marginal-cost
driven also means that they can easily adapt to relatively slow
changes in the network topology or traffic patterns.  For if channel
gains and/or traffic input rates change, then the relevant marginal
costs change accordingly, and the node iterations naturally adapt to
the new network conditions by responding to the new marginal costs.
The adaptability of the algorithms to changing network conditions is
confirmed in numerical experiments presented in
Section~\ref{sec:Simulation_change}.



\section{\label{sec:GeneralizationAndRelaxation}Refinements and Generalizations}\vspace{3mm}

In this section, we introduce a number of refinements and
generalizations to improve the applicability and utility of our
analytical framework and proposed algorithms. Specifically, we
consider three main issues.  First, we present a refinement of the
power allocation algorithm for CDMA networks with single-user
decoding by relaxing the high-SINR assumption
in~\eqref{eq:HighSINRCapacity}.  This assumption has thus far
limited the range of feasible controls for the power allocation and
power control algorithms.  To address this problem, we introduce a
heuristic two-stage network optimization scheme which significantly
enlarges the range of control possibilities. Next, we generalize the
SINR-dependent network model to analyze wireless networks operating
with general physical-layer coding schemes. Instead of assuming
concave capacity functions dependent on the links' SINR, we assume
link capacities are given by a general convex achievable rate
region. We then characterize the optimality conditions for the JOCR
problem given a general convex rate region. Finally, we relax the
requirement that the link cost functions are jointly convex in the
link capacities and link flow rates.  This joint convexity
assumption was needed to prove that the necessary conditions for
global optimality are also sufficient. We show that if cost
functions satisfy the less stringent requirement of strict {\em
quasiconvexity}, then solutions satisfying the necessary conditions
for optimality still have the desirable property of being {\em
Pareto optimal} when the underlying capacity region is strictly
convex.

\subsection{Refined Power Allocation and Two-Stage Network Optimization}

\label{sec:MicroMacro}


Our formulation of the joint power control and routing problem in
\eqref{eq:JOPR}-\eqref{eq:CapacityFormulaApprox} rests on the
crucial condition~\eqref{eq:ConcaveCapacity} on the capacity
function. Such an assumption implies that $\lim_{x\to 0^+}C''(x) =
-\infty$ since by monotonicity $\lim_{x\to 0^+}C'(x) > 0$. However,
this yields the rather disturbing result that $\lim_{x\to 0^+}C'(x)
= \infty$ and $\lim_{x\to 0^+}C(x) = -\infty$. The approximate
information-theoretic capacity~\eqref{eq:HighSINRCapacity} and the
M-QAM capacity~\eqref{eq:MQAMCapacity} with error probability
constraint satisfy~\eqref{eq:ConcaveCapacity}, but are both based on
the high-SINR approximation. Indeed, since CDMA networks typically
do have high \emph{per symbol} SINR due to the large processing gain
$K$, $C = \log(K \cdot SINR)$ have been extensively used as a
reasonable approximate capacity function for CDMA networks in
previous literature~\cite{paper:Chi04,paper:JXB03}. Outside of the
high-SINR regime, however, $C = \log(K \cdot SINR)$ becomes too
inaccurate to be applicable because, for instance, it gives $C < 0$
when $SINR < 1 / K$ and $C = -\infty$ when $SINR = 0$. Thus,
adopting $C = \log(K \cdot SINR)$ as the capacity function
significantly restricts the optimization of transmission powers and
traffic flows.\footnote{Note that if the network running the RT, PA,
and PC algorithms described above starts with a control
configuration with {\em finite} cost, then the capacity of each link
$(i,j)$ (under the high-SINR assumption) must be positive, implying
that $SINR_{ij} > 1/K$. Since the algorithms reduce the total
network cost with each iteration, the condition $SINR_{ij} > 1/K$
continues to hold with each iteration.  Moreover, since the
high-SINR assumption {\em under}estimates the actual link capacity,
the power control and routing configurations resulting from RT, PA,
and PC are always {\em feasible.}}

%

{Ideally, instead of $\log(K\cdot SINR)$, we would use the precise
capacity function $C = \log(1 + K \cdot SINR)$. Note that the latter
function does not satisfy~\eqref{eq:ConcaveCapacity}, and does not
lead to a convex JOPR problem in the original framework of
Section~\ref{sec:OptimizationSpaces}. However, we show that if the
total powers of individual nodes $\{P_i\}$ (or equivalently
$\{\gamma_i\}$) are held {\em fixed}, the precise capacity function
does give rise to a {\em convex} optimization problem in typical
CDMA networks. In other words, the JOPR problem involving only
routing and power allocation is convex in the optimization variables
$\{\phi_{ij}(w)\}$ and $\{\eta_{ij}\}$ when the link capacities are
given by $C = \log(1 + K \cdot SINR)$. We call this revised problem
the {\em Jointly Optimal Power Allocation and Routing} (JOPAR)
problem. 

\subsubsection{Concavity of the Precise Capacity Function}

{Since the change of link capacity functions does not alter the convexity of the
objective function with respect to the flow variables, we need only verify that the
objective function is jointly convex in the power allocation variables $\{\eta_{ij}\}$.
This is equivalent to showing that each link capacity function
\begin{equation}\label{eq:CapacityPrecise} C_{ij} = \log \left(1 + \displaystyle\frac{KG_{ij}P_i \eta_{ij}}{
G_{ij} P_i ( 1 - \eta_{ij}) + \sum\limits_{m \ne i}{G_{mj}P_m} + N_j }\right).
\end{equation} is concave in
$\eta_{ij}$.}

\vspace{0.1in}\begin{lemma}\label{lma:MicroRegionConvex} Link capacity $C_{ij}$ given by~\eqref{eq:CapacityPrecise} is
concave in $\eta_{ij}$ if the following interference-limited condition holds:
\begin{equation}\label{eq:IntfLimitCond}
KG_{ij} P_{ij}  \le \left( {K - 2} \right)IN_{ij}.
\end{equation}
\end{lemma}\vspace{0.1in}

{Note that the condition \eqref{eq:IntfLimitCond} is almost always satisfied in CDMA
systems, where interference level $IN_{ij}$ is usually higher than that of the received
signal power $G_{ij} P_{ij}$ by several orders of magnitude \cite{book:TV04}.}
\vspace{0.1in}

\textit{Proof of Lemma~\ref{lma:MicroRegionConvex}:} {Differentiating the RHS of
\eqref{eq:CapacityPrecise} twice with respect to $\eta_{ij}$
\begin{displaymath}
\frac{d^2 C_{ij}}{d \eta_{ij}^2} = P_i^2 \left\{ - \left[\frac{(K - 1) G_{ij} }{IN_{ij} + KG_{ij} P_{ij}}\right]^2 +
\left[\frac{ G_{ij}}{IN_{ij}}\right]^2\right\}.
\end{displaymath}
Using \eqref{eq:IntfLimitCond}, we have
\begin{displaymath}
\frac{d^2 C_{ij}}{d\eta_{ij}^2} \le P_i^2 \left\{ -\left[\frac{(K - 1) G_{ij}}{IN_{ij} + \left( {K - 2}\right)
IN_{ij}}\right]^2 + \left[\frac{ G_{ij}}{IN_{ij}}\right]^2 \right\} = 0,
\end{displaymath}
which implies that $C_{ij}$ is concave in $\eta_{ij}$.} \qed

\subsubsection{Power Allocation and Routing for JOPAR Problem}

{The JOPAR problem holds $\{\gamma_i\}$ fixed, so its solution is obtained only through
varying $\{\phi_{ij}(w)\}$ (routing) and $\{\eta_{ij}\}$ (power allocation). In
particular, the routing scheme is unchanged from that for the original
problem~\eqref{eq:JOPR}. On the other hand, the marginal power allocation cost needs to
be revised according to~\eqref{eq:CapacityPrecise} as
\begin{equation}\label{eq:MicroPowerMargCost}
\delta\eta_{ij} = \frac{\partial D_{ij}}{\partial C_{ij}}\left(\frac{(K - 1) G_{ij}}{K G_{ij} P_i \eta_{ij} + IN_{ij}}
+ \frac{ G_{ij}}{IN_{ij}}\right), \; j \in {\cal O}(i).
\end{equation}
With $\{\delta\phi_{ij}(w)\}$ and $\{\delta\eta_{ij}\}$ given
by~\eqref{eq:MarginalFlowCost} and \eqref{eq:MicroPowerMargCost}, the optimality
conditions for the JOPAR problem are stated as in Theorem~\ref{thm:OptimalityCondition}
with \eqref{eq:PowerOptCond2} and \eqref{eq:PowerOptCond3} removed.}

{We now specify the power allocation algorithm $(PA)$ for the JOPAR
problem. It retains the same scaled gradient projection form as
in~\eqref{eq:GeneralPAUpdate} but with the scaling matrix $Q_i^k$
given differently as follows.}

\vspace{0.1in}\begin{lemma}\label{lma:MicroHessianBnd} If the
current local cost is $\sum_{j \in {\cal O}(i)}D_{ij}^k = D_i^k$,
then at the current iteration of the $PA$ algorithm
in~\eqref{eq:GeneralPAUpdate} (with revised $(\delta \eta_{ij})$
given by~\eqref{eq:MicroPowerMargCost}) and for all $\lambda \in
[0,1]$, the Hessian matrix $H^{k,\lambda}_{\bs\eta_i}=\left.\nabla^2
D(\bs\eta_i)\right|_{\bs\eta_i = \lambda\bs\eta_i^k +
(1-\lambda)\bs\eta_i^{k+1}}$ is upper bounded by the diagonal matrix
\[
\bar Q_i^k = {\rm diag}\left\{\left(\left[\bar B^k_{ij}(D_i^k)K^2 - \underline B^k_{ij}(D_i^k)\left( {K-1} \right)^2
\right] (NR_{ij})^2\right)_{j \in {\cal O}(i)}\right\},
\]
where \be \bar B^k_{ij}(D_i^k) \equiv \max_{C_{ij}:D_{ij}(C_{ij},F^k_{ij}) \le D_i^k}\frac{\partial^2 D_{ij}}{\partial
C_{ij}^2}, \label{eq:barBij} \ee \be \underline B^k_{ij}(D_i^k) \equiv \min_{C_{ij}:D_{ij}(C_{ij},F^k_{ij}) \le
D_i^k}\frac{\partial D_{ij}}{\partial C_{ij}}, \label{eq:underlineBij} \ee and \be NR_{ij} \equiv \frac{ G_{ij}
P_i}{\sum_{m \ne i} {G_{mj} P_m} + N_j}. \label{eq:NRij} \ee
\end{lemma}\vspace{0.1in}

{The proof of the lemma is in Appendix~\ref{app:MicroEtaHessian}. Accordingly, the
stepsize for the BPA algorithm~\eqref{eq:BasicPA} can be chosen as
\be\label{eq:BasicRevisedPAStepsizeRange} \beta_i^k = 2 P_i^2 \left[|{\cal O}(i)|\max_{j
\in {\cal O}(i)} \left[\bar B^k_{ij}(D_i^k)K^2 - \underline B^k_{ij}(D_i^k)\left( {K-1}
\right)^2 \right] (NR_{ij})^2 \right]^{-1}. \ee One can also apply the GPA
algorithm~\eqref{eq:GeneralPAUpdate} for the JOPAR problem. In this case, the scaling
matrix is given by
\[
Q_i^k = \frac{\bar Q_i^k}{2 P_i}.
\]
Such a choice of $\beta_i^k$ and $Q_i^k$ guarantees that any
iteration of the BPA and GPA algorithms strictly reduces the network
cost unless condition \eqref{eq:PowerOptCond1} is satisfied. As a
result, the refined power allocation algorithm and the routing
algorithm can converge to an optimal solution of the JOPAR problem
from any initial configuration of $\{\phi_{ij}(w)\}$ and
$\{\eta_{ij}\}$.}

\subsubsection{Heuristic Two-Stage Network Optimization}

{The refined power allocation technique based on the precise
capacity formula allows us to adjust link powers over their full
range from zero to the total power of their respective
transmitters.\footnote{More precisely, in order to keep the link
cost finite, the refined power allocation algorithm only allows one
to reduce link powers arbitrarily close to zero.} This fine-tuning
capability, however, comes at the expense of fixing the total power
of nodes. Should the node powers $(P_i)$ be variable, the capacity
function $\log(1+K\cdot SINR(\bs P))$ would no longer be concave in
link power variables. Although the power control algorithm in
Section~\ref{sec:ConvergenceOfAlgorithms} is built on the high-SINR
approximation, in practice it can be applied in conjunction with the
routing algorithm and the refined power allocation algorithm
developed above.}

To carry out the overall task of routing and power adjustment, we
let the nodes iterate between a routing/power allocation stage and a
power control stage. In the routing/power allocation stage, nodes
adjust their routing variables $\phi_{ij}(w)$ and power allocation
variables $\eta_{ij}$ as in the JOPAR problem discussed above
according to the refined $PA$ algorithm while {\em holding the total
transmission power $P_i$ fixed}, evaluating link capacities by the
precise $\log(1+K\cdot SINR(\bs P))$ formula.  As pointed above,
this routing/power allocation stage can asymptotically achieve the
optimal set of $(\eta_{ij})$ and $(\phi_{ij}(w))$ for the given
total powers $(P_i)$.

To further (strictly) reduce the total cost, one can switch to the
power control stage, where total power $P_i$'s are adjusted by the
power control algorithm~\eqref{eq:BasicPC} while {\em holding the
routing variables $\phi_{ij}(w)$ and power allocation variables
$\eta_{ij}$ fixed}. By using the approximate $\log(K\cdot SINR(\bs
P))$ formula in the power control stage, the total cost is convex in
the power control variables $(\gamma_i)$. Power control algorithms
thus can converge to the optimal total powers under the fixed
routing $(\phi_{ij}(w))$ and power allocation $(\eta_{ij})$.

Heuristically, one can then iterate between the routing/power
allocation and power control stages to arrive at a network
configuration that is approximately optimal.


\subsection{General Capacity Regions}

\label{sec:GenCapRegion}

Up to this point, we have assumed that link capacities are functionally determined by the
links' SINR.  Under individual power constraints \eqref{eq:TotalPowerConstraint} and
assumption~\eqref{eq:ConcaveCapacity}, the achievable link capacities were shown to
constitute a convex set. In order to place our analysis and algorithms in a broader
setting where more general coding/modulation schemes are applied, we now consider the
general JOCR problem~\eqref{eq:JOCR} where the achievable rate region ${\cal C}$ is any
convex set in the positive orthant $\mathbb R_+^{|{\cal E}|}$.  The convexity assumption
is reasonable since any convex combination of a pair of feasible link capacity vectors
can at least be achieved by time-sharing or frequency-sharing.

The following theorem characterizes the optimality conditions for
the JOCR problem with a general convex capacity region.

\vspace{0.1in}\begin{theorem}\label{thm:GeneralOptCond} Assume that the cost functions $D_{ij}(C_{ij},F_{ij})$ satisfy
\eqref{eq:CostFunctionIdentity} and assume that ${\cal C}$ is convex. Then, for a feasible set of routing and capacity
allocations $(\phi_{ij}(w))_{w \in {\cal W}, (i,j) \in {\cal E}}$ and $(C_{ij})_{(i,j) \in {\cal E}}$ to be a solution
of JOCR \eqref{eq:JOCR}, the following conditions are necessary. For all $i \in {\cal N}$ and $w \in {\cal W}$ such
that $t_i(w) > 0$, there exists a constant $\lambda_i(w)$ for which
\begin{equation}\label{eq:GeneralRoutingOptCond}
\begin{array}{ll}
\vspace{1mm} \delta\phi_{ij}(w) = \lambda_i(w), \; &\textrm{if}~\phi_{ij}(w) > 0,
\\
\vspace{1mm} \delta\phi_{ij}(w) \ge \lambda_i(w), &\textrm{if}~\phi_{ij}(w) = 0.
\end{array}
\end{equation}
For all feasible $(\Delta C_{ij})_{(i,j) \in {\cal E}}$ at $(C_{ij})_{(i,j) \in {\cal E}}$,
\begin{equation}\label{eq:GeneralCapacityOptCond}
\sum_{(i,j) \in {\cal E}}{\frac{\partial D_{ij}}{\partial C_{ij}}(C_{ik},F_{ik}) \cdot \Delta C_{ij}} \ge 0,
\end{equation}
where an incremental direction $(\Delta C_{ij})_{(i,j) \in {\cal E}}$ at $(C_{ij})_{(i,j) \in {\cal E}}$ is said to be
feasible if there exists $\bar\delta > 0$ such that $(C_{ij} + \delta \cdot \Delta C_{ij})_{(i,j) \in {\cal E}} \in
{\cal C}$ for any $\delta \in (0,\bar\delta)$.

If $D_{ij}(C_{ij},F_{ij})$ is jointly convex in $(C_{ij},F_{ij})$, the above conditions are also sufficient when
\eqref{eq:GeneralRoutingOptCond} holds for all $i \in {\cal N}$ and $w \in {\cal W}$ whether $t_i(w) > 0$ or not.
Furthermore, the optimal $(C_{ij}^*)_{(i,j) \in {\cal E}}$ is unique if ${\cal C}$ is strictly convex. If, in addition,
$D_{ij}(C_{ij},F_{ij})$ is strictly convex in $F_{ij}$, then the optimal link flows $(F_{ij}^*)_{(i,j) \in {\cal E}}$
are unique as well.
\end{theorem}\vspace{0.1in}

\textit{Proof:} The necessity and sufficiency statements can be proved by following the
same argument used for proving Theorem \ref{thm:OptimalityCondition}. Thus, we do not
repeat it here. We show only the uniqueness of the optimal $(C_{ij})$ and $(F_{ij})$
under the respective assumptions.

Suppose on the contrary, there are two distinct optimal solutions $\{(C_{ij}^0),~(F_{ij}^0)\}$ and
$\{(C_{ij}^1),~(F_{ij}^1)\}$ such that $(C_{ij}^0) \ne (C_{ij}^1)$ and their common minimal cost is $D^*$. Consider the
total cost resulting from $\{(C_{ij}^\lambda),~(F_{ij}^\lambda)\}$, where $C_{ij}^\lambda = \lambda C_{ij}^0 +
(1-\lambda) C_{ij}^1$, $F_{ij}^\lambda = \lambda F_{ij}^0 + (1-\lambda) F_{ij}^1$ for all $(i,j)\in {\cal E}$ and for
some $\lambda \in (0,1)$.

By the joint convexity of $D_{ij}(\cdot,\cdot)$, we have for all $(i,j)\in {\cal E}$,
\[
D_{ij}(C_{ij}^\lambda, F_{ij}^\lambda) \le \lambda D_{ij}(C_{ij}^0, F_{ij}^0) + (1-\lambda) D_{ij}(C_{ij}^1, F_{ij}^1).
\]
If ${\cal C}$ is strictly convex and $\{C_{ij}^0\} \ne \{C_{ij}^1\}$, there must exist $\{\bar C_{ij}^\lambda\} \in
{\cal C}$ such that
\[
\bar C_{ij}^\lambda \ge C_{ij}^\lambda, \; \forall (i,j)\in {\cal E}
\]
with at least one inequality being strict. Without loss of generality assume $\bar C_{mn}^\lambda > C_{mn}^\lambda$.
Using the fact that $\frac{\partial D_{ij}}{\partial C_{ij}} < 0$ for all $(i,j)$, we have $D_{ij}(\bar C_{ij}^\lambda,
F_{ij}^\lambda) \le D_{ij}(C_{ij}^\lambda, F_{ij}^\lambda)$ and in particular $D_{mn}(\bar C_{mn}^\lambda,
F_{mn}^\lambda) < D_{mn}(C_{mn}^\lambda, F_{mn}^\lambda)$. Therefore, summing over all links,
\[
\sum_{(i,j) \in {\cal E}} D_{ij}(\bar C_{ij}^\lambda, F_{ij}^\lambda) < \sum_{(i,j) \in {\cal E}}
D_{ij}(C_{ij}^\lambda, F_{ij}^\lambda) \le \sum_{(i,j) \in {\cal E}} \lambda D_{ij}(C_{ij}^0, F_{ij}^0) + (1-\lambda)
D_{ij}(C_{ij}^1, F_{ij}^1) = D^*.
\]
Since $\{(\bar C_{ij}^\lambda),~(F_{ij}^\lambda)\}$ is feasible, the above inequality contradicts the optimality of
$D^*$. When $D_{ij}(C_{ij},\cdot)$ is strictly convex in $F_{ij}$, a similar contradiction arises if the optimal
$(F_{ij})$'s are not unique. Thus the proof is complete. \qed

\vspace{3mm}\subsection{Quasiconvex Cost Functions and Pareto Optimality}

\label{sec:QuasiConvex}

\vspace{3mm}After considering general convex capacity regions, we
turn our attention to the network cost measures.  The sufficiency of
conditions
\eqref{eq:GeneralRoutingOptCond}-\eqref{eq:GeneralCapacityOptCond}
for global optimality depends on link cost functions
$D_{ij}(C_{ij},F_{ij})$ being jointly convex in $(C_{ij},F_{ij})$.
Without the joint convexity assumption,
inequality~\eqref{eq:ObjectiveDiff} is no longer valid, and the
sufficiency parts of Theorem~\ref{thm:OptimalityCondition} and
Theorem~\ref{thm:GeneralOptCond} do not hold. On the other hand, our
initial assumptions in~\eqref{eq:CostFunctionIdentity} regarding the
cost functions do not imply $D_{ij}(C_{ij},F_{ij})$ is jointly
convex. In particular, the often-used cost function
$\frac{F_{ij}}{C_{ij}-F_{ij}}$ is not jointly convex. We show in the
following, however, that if the cost functions satisfy the less
stringent requirement of {\em strict quasiconvexity}, then solutions
satisfying~\eqref{eq:GeneralRoutingOptCond}-\eqref{eq:GeneralCapacityOptCond}
are in fact {\em Pareto optimal}.

In the subsequent analysis, we assume that for all $(i,j) \in {\cal E}$,
$D_{ij}(C_{ij},F_{ij})$ is twice continuously differentiable on ${\cal X} = \{(C_{ij},
F_{ij}): 0 \leq F_{ij} < C_{ij}\}$, and satisfies~\eqref{eq:CostFunctionIdentity}.
Furthermore, we assume that $D_{ij}(C_{ij},F_{ij})$ is \emph{strictly quasiconvex}, i.e.
if
\begin{equation}
D_{ij}(C_{ij}^1,F_{ij}^1) \le D_{ij}(C_{ij}^2,F_{ij}^2),
\label{eq:quasi1}
\end{equation}
then
\begin{equation}
D_{ij}(\lambda C_{ij}^1 + (1-\lambda)C_{ij}^2,\lambda F_{ij}^1 +
(1-\lambda)F_{ij}^2) \le D_{ij}(C_{ij}^2,F_{ij}^2), \; \forall
\lambda \in (0,1), \label{eq:quasi2}
\end{equation}
with strict inequality in~\eqref{eq:quasi1} implying strict
inequality in~\eqref{eq:quasi2}.  It is easily verified that the
cost functions $ {F_{ij}}/(C_{ij}-F_{ij})$ and $
{1}/(C_{ij}-F_{ij})$ are both strictly quasiconvex.

We consider the general JOCR problem \eqref{eq:JOCR} where ${\cal C}\in \mathbb
R_+^{|{\cal E}|}$ is \emph{strictly convex}. Due to
assumption~\eqref{eq:CostFunctionIdentity}, for a fixed capacity allocation $(C_{ij})
_{(i,j)\in {\cal E}}$, \eqref{eq:JOCR} is a convex optimization problem with respect to
$(F_{ij}(w))_{(i,j)\in {\cal E}}$. Hence, any feasible flow distribution $( F_{ij}^*)
_{(i,j)\in {\cal E}}$ satisfying \eqref{eq:GeneralRoutingOptCond} also satisfies
\begin{equation}\label{eq:FlowMinimizer}
\sum_{(i,j)\in {\cal E}} {D_{ij}\left( C_{ij}, F_{ij}^*\right)} = \min_{( F_{ij}) \textrm{ feasible}} {\sum_{(i,j)\in
{\cal E}} {D_{ij}\left( C_{ij}, F_{ij}\right)}}.
\end{equation}

On the other hand, given any feasible routing configuration, if condition \eqref{eq:GeneralCapacityOptCond} holds at
capacity allocation $(C_{ij}^*)_{(i,j)\in {\cal E}}$, then it follows that
\begin{equation}\label{eq:CapacityMinimizer}
\sum_{(i,j)\in {\cal E}} {D_{ij}\left( C_{ij}^*, F_{ij}\right)} = \min_{(C_{ij}) \in {\cal C}} {\sum_{(i,j)\in {\cal
E}} {D_{ij}\left( C_{ij}, F_{ij}\right)}}.
\end{equation}

Under the capacity model in Section~\ref{subsec:CapacityModel}, the algorithms proposed in
Section~\ref{sec:ConvergenceOfAlgorithms} have been shown to drive any initial routing and capacity configuration to a
limiting $\left\{ (C_{ij}^*)_{(i,j)\in {\cal E}}, (F_{ij}^*)_{(i,j)\in {\cal E}}\right\} $ such that the condition
\eqref{eq:GeneralRoutingOptCond} is satisfied at $( F_{ij}^* )$ given $(C_{ij}^*)$, and $( C_{ij}^*)$ satisfies
\eqref{eq:GeneralCapacityOptCond} given $( F_{ij}^*)$. Under the more general convex-capacity-region model, suppose we
have algorithms that also can drive the flow and capacity configuration to a limit $\left\{ (C_{ij}^*),
(F_{ij}^*)\right\} $ such that the conditions \eqref{eq:GeneralRoutingOptCond}-\eqref{eq:GeneralCapacityOptCond} hold
simultaneously. We are then interested in the question: to what extent can optimality be inferred from such a limit
point? Although global optimality cannot be ascertained, we have the following \emph{Pareto optimal} property.

\vspace{0.1in}\begin{theorem}\label{thm:ParetoOptimality} Assume that the capacity region
${\cal C}$ is strictly convex and the link cost function is strictly quasiconvex. If a
pair of feasible capacity and flow rate allocations $\left\{ (C_{ij}^*), (F_{ij}^*)
\right\}$ satisfies conditions \eqref{eq:GeneralRoutingOptCond} and
\eqref{eq:GeneralCapacityOptCond} simultaneously, then the vector of link costs $\bs
D\left( \bs C^*, \bs F^*\right) \triangleq \left( D_{ij}\left( C_{ij}^*, F_{ij}^*\right)
\right) _{(i,j)\in {\cal E}}$ is Pareto optimal, i.e. there does not exist another pair
of feasible allocations $ \{( C_{ij}^{\#}), (F_{ij}^{\#}) \}$ such that
\begin{displaymath}
D_{ij} ( C_{ij}^{\#}, F_{ij}^{\#} ) \le D_{ij}\left( C_{ij}^{*}, F_{ij}^{*}\right), \; \forall (i,j)\in {\cal E},
\end{displaymath}
with at least one inequality being strict.
\end{theorem}\vspace{0.1in}

Assuming the cost function $D_{ij} = F_{ij}/(C_{ij}-F_{ij})$,
Theorem~\ref{thm:ParetoOptimality} can be taken to mean that at the
(Pareto) optimal point, the average number of packets cannot be
strictly reduced on one link without it being increased on another.
\vspace{0.1in}

\textit{Proof of Theorem~\ref{thm:ParetoOptimality}:} Suppose on the
contrary $\bs D\left( \bs C^\#, \bs F^\#\right)$ Pareto dominates
$\bs D\left( \bs C^*, \bs F^*\right)$. Without loss of generality,
assume \[ D_{mn}\left( C_{mn}^{\#}, F_{mn}^{\#}\right) <
D_{mn}\left( C_{mn}^{*}, F_{mn}^{*}\right).
\]
Because both $\bs C^\#$ and $\bs C^*$ belong to ${\cal C}$, and ${\cal C}$ is strictly
convex, $\bs C^\lambda = \lambda \bs C^* + (1-\lambda)\bs C^\#$ is achievable for all
$\lambda \in [0,1]$. Moreover, it can be deduced that $\bs C^\# \ne \bs C^*$, since
otherwise we must have $\bs F^\# \ne \bs F^*$, and Pareto domination would imply
$\sum_{(i,j)\in {\cal E}}{D_{ij}\left( C_{ij}^{*}, F_{ij}^{\#}\right)} < \sum_{(i,j)\in
{\cal E}}{D_{ij}\left( C_{ij}^{*}, F_{ij}^{*}\right)}$, hence contradicting identity
\eqref{eq:FlowMinimizer}. Therefore, we conclude that $\bs C^\lambda$ is in the interior
of ${\cal C}$ for any $\lambda \in (0,1)$. Using the same reasoning, we can assert that
$\bs F^\# \ne \bs F^*$, and that $\bs F^\lambda = \lambda \bs F^* + (1-\lambda)\bs F^\#$
is feasible for any $\lambda \in [0,1]$ simply by the linearity of the flow conservation
constraint.

As a consequence of $D_{ij}$ being strictly quasiconvex, $\bs D\left(\bs C^\lambda, \bs
F^\lambda \right)$ Pareto dominates $\bs D\left( \bs C^*, \bs F^* \right)$ as well for
any $\lambda \in (0,1)$, since $D_{mn}\left( C_{mn}^{\lambda}, F_{mn}^{\lambda}\right) <
D_{mn}\left( C_{mn}^{*}, F_{mn}^{*}\right)$ and $D_{ij}\left( C_{ij}^{\lambda},
F_{ij}^{\lambda}\right) \le D_{ij}\left( C_{ij}^{*}, F_{ij}^{*}\right)$, for $(i,j) \ne
(m,n)$. Summing up all the terms on LHS and RHS, we have
\begin{equation}\label{eq:ParetoProof1}
\displaystyle\sum_{(i,j)\in {\cal E}}{D_{ij}\left(
C_{ij}^{\lambda}, F_{ij}^{\lambda}\right)} <
\displaystyle\sum_{(i,j)\in {\cal E}}{D_{ij}\left( C_{ij}^{*},
F_{ij}^{*}\right)}, \; \forall \lambda \in (0,1).
\end{equation}

By optimality condition \eqref{eq:GeneralCapacityOptCond} and the
fact that $\bs C^\lambda$ is in the interior of ${\cal C}$ for any
$\lambda \in (0,1)$, we have
\begin{equation}\label{eq:ParetoProof2}
\begin{array}{ll}
\vspace{1mm} & \displaystyle\sum_{(i,j)\in {\cal
E}}{\frac{\partial D_{ij}}{\partial C_{ij}}\left( C_{ij}^{*},
F_{ij}^{*}\right) \cdot \left( C_{ij}^\# - C_{ij}^* \right)} \\
\vspace{1mm} = &
\displaystyle\frac{1}{1-\lambda}\displaystyle\sum_{(i,j)\in {\cal
E}}{\frac{\partial D_{ij}}{\partial C_{ij}}\left( C_{ij}^{*},
F_{ij}^{*}\right) \cdot \left( C_{ij}^\lambda - C_{ij}^* \right)} \\
\vspace{1mm}
> & \displaystyle\frac{1}{1-\lambda}\displaystyle\sum_{(i,j)\in
{\cal E}}{\frac{\partial D_{ij}}{\partial C_{ij}}\left(
C_{ij}^{*}, F_{ij}^{*}\right) \cdot \left( \bar C_{ij}^\lambda -
C_{ij}^* \right)} \ge 0, \; \forall \lambda \in (0,1),
\end{array}
\end{equation}
where $( \bar C_{ij}^\lambda )_{(i,j)\in {\cal E}}$ is some capacity vector that strictly dominates $( C_{ij}^\lambda
)_{(i,j)\in {\cal E}}$.

Since $D_{ij}$ is twice continuously differentiable, there exists $\varepsilon > 0$ such that for all $\lambda \in
[1-\varepsilon, 1)$,
\begin{displaymath}
\displaystyle\sum_{(i,j)\in {\cal E}}{\frac{\partial
D_{ij}}{\partial C_{ij}}\left( C_{ij}^{*}, F_{ij}^{\lambda}\right)
\cdot \left( C_{ij}^\lambda - C_{ij}^* \right)} \ge 0,
\end{displaymath}
which, combined with the convexity of $D_{ij}\left( \cdot, F_{ij}^\lambda \right)$, implies
\[
\displaystyle\sum_{(i,j)\in {\cal E}}{D_{ij}\left( C_{ij}^{*}, F_{ij}^{\lambda}\right)}  \le
\displaystyle\sum_{(i,j)\in {\cal E}}{D_{ij}\left( C_{ij}^{\lambda}, F_{ij}^{\lambda}\right)} <
\displaystyle\sum_{(i,j)\in {\cal E}}{D_{ij}\left( C_{ij}^{*}, F_{ij}^{*}\right)},
\]
where the second inequality comes from \eqref{eq:ParetoProof1}.
But this conclusion clearly contradicts \eqref{eq:FlowMinimizer}.
Thus, the claim is proved. \qed


\section{\label{sec:ElasticDemand}Congestion Control}

Thus far, we have focused on developing optimal power control and routing algorithms for
given fixed user traffic demands. There are many situations, however, where the resulting
network delay cost is excessive for given user demands even with optimal power control
and routing.  In these cases, congestion control must be used to limit traffic input into
the network.  In this section, we extend our analytical framework to consider congestion
control for sessions with elastic traffic demands.  We show that congestion control can
be seamlessly incorporated into our framework, in the sense that the problem of jointly
optimal power control, routing, and congestion control can always be converted into a
problem involving only power control and routing.

\subsection{User Utility, Network Cost, and Congestion Pricing}

For a given session $w$, let the utility level associated with an admitted rate of $r_w$ be $U_w(r_w)$. We consider
maximizing the \emph{aggregate session utility minus the total network cost} \cite{paper:Kel97}, i.e.
\be\label{eq:UtltyMinusCost} \textrm{maximize} \sum_{w \in {\cal W}} U_w(r_w) - \sum_{(i,j) \in {\cal E}}
D_{ij}(C_{ij},F_{ij}). \ee


We make the reasonable assumption that each session $w$ has a maximum desired service rate $\bar r_w$.  The session
utility $U_w(\cdot)$ is defined over the interval $[0,\bar r_w]$, where it is assumed to be twice continuously
differentiable, strictly increasing, and concave.  Taking the approach of~\cite{book:BG92}, we define the {\em overflow
rate} $F_{wb} \triangleq \bar r_w - r_w \geq 0$ for a given admitted rate $r_w \leq \bar{r}_w$. Thus, at each source
node $i = O(w)$, we have
\begin{equation}\label{eq:SourceFlowCons}
\sum_{j \in {\cal O}(i)} {f_{ij}(w)} + F_{wb} = \bar r_w.
\end{equation}

Let $B_w(F_{wb}) \equiv U_w(\bar r_w) - U_w(r_w)$ denote the utility loss for session $w$ resulting from having a rate
of $F_{wb}$ rejected from the network.  Equivalently, if we imagine that the blocked flow $F_{wb}$ is routed on a {\em
virtual overflow link} directly from the source to the destination \cite{book:BG92}, then $B_w(F_{wb})$ can simply be
interpreted as the cost incurred on that virtual link when its flow rate is $F_{wb}$. Moreover, as defined,
$B_w(F_{wb})$ is strictly increasing, twice continuously differentiable, and convex in $F_{wb}$ on $[0, \bar{r}_w]$.
Thus, the dependence of $B_w(F_{wb})$ on $F_{wb}$ is the same as the dependence of the cost functions $D_{ij}(C_{ij},
F_{ij})$ of real links $(i,j)$ on the flow $F_{ij}$.  Unlike $D_{ij}(C_{ij}, F_{ij})$, however, $B_w$ has no explicit
dependence on a capacity parameter.\footnote{If we assume that $U_w(0) = -\infty$, so that there is an infinite penalty
for admitting zero session $w$ traffic, then $B_w(\bar{r}_w) =\infty$, and $\bar{r}_w$ could be taken as the (fixed)
``capacity" of the overflow link.} A virtual network including an overflow link is illustrated in Figure
\ref{fig:OverflowLink}, where the overflow link $wb$ is marked by a dashed arrow.

\begin{figure}
\begin{center}
\includegraphics[width = 10cm]{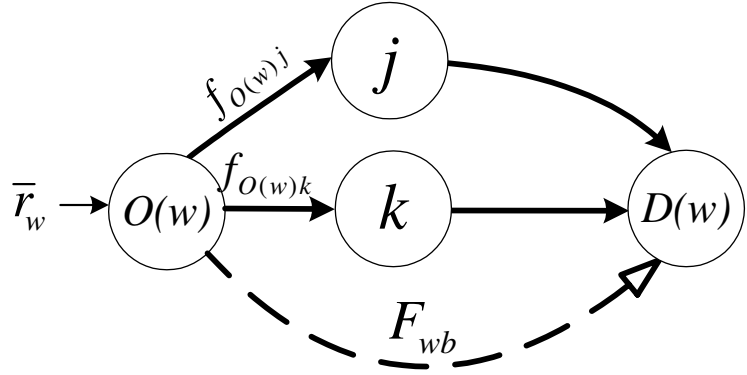}
\caption{Virtual Network with Overflow Link}\label{fig:OverflowLink}
\end{center}
\end{figure}

Accordingly, the objective in~\eqref{eq:UtltyMinusCost} can now be written as
\begin{equation}\label{eq:SumOfPayoffsTrans}
\sum_{w \in {\cal W}} U_w(r_w) - \sum_{(i,j) \in {\cal E}} D_{ij}(C_{ij},F_{ij}) = \sum_{w \in {\cal W}} U_w(\bar r_w)
- \sum_{w \in {\cal W}} B_w(F_{wb}) - \sum_{(i,j) \in {\cal E}} D_{ij}(C_{ij},F_{ij}).
\end{equation}
Since $\sum_{w \in {\cal W}} U_w(\bar r_w)$ is a constant,~\eqref{eq:UtltyMinusCost} is equivalent to
\begin{equation}\label{eq:FlowCtrlObj}
\textrm{minimize}~\sum_{(i,j) \in {\cal E}} D_{ij}(C_{ij},F_{ij}) + \sum_{w \in {\cal W}} B_w(F_{wb}).
\end{equation}
Note that~\eqref{eq:FlowCtrlObj} has the same form as~\eqref{eq:JOCR}, except for the lack of dependence of
$B_w(F_{wb})$ on a capacity parameter.   Thus, the problem of jointly optimal power control, routing, and congestion
control in a wireless network is equivalent to a problem involving only power control and routing in a virtual wireless
network with the addition of the virtual overflow links.

\vspace{3mm}\subsection{Optimal Distributed Power Control,
Routing, and Congestion Control}\vspace{3mm}

We now turn to developing distributed network algorithms to solve the jointly optimal power control, routing, and
congestion control problem in~\eqref{eq:UtltyMinusCost}.  We show that the network algorithms developed in
Section~\ref{sec:ConvergenceOfAlgorithms} can readily be adapted to deal with the new challenge of congestion control.
Thus, seemingly disparate network functionalities at the physical, medium access, network, and transport layers of the
traditional OSI hierarchy are naturally combined into a common framework.


To specify the distributed algorithms, we continue to use the routing, power allocation, and power control variables,
except for a modification of the definition of the routing variables $\boldsymbol\phi_i(w)$ at $i = O(w)$, $w \in {\cal
W}$. Define
\begin{equation}\label{eq:SourceRTVarRedef}
\begin{array}{l}
\vspace{1mm} t_i(w) \triangleq \bar r_w,
\\
\vspace{1mm} \displaystyle{\phi_{ij}(w) \triangleq \frac{f_{ij}(w)}{t_i(w)}, \; \forall j \in {\cal O}(i)},
\\
\vspace{1mm} \displaystyle{\phi_{wb} \triangleq \frac{F_{wb}}{t_i(w)}}.
\\
\end{array}
\end{equation}
The new routing variables are subject to the simplex constraint
\[
\phi_{ij} (w) \ge 0, \; \phi_{wb} \ge 0, \; \sum_{j \in {\cal O}(i)} {\phi_{ij}(w)} + \phi_{wb} = 1.
\]

We now state the Jointly Optimal Power Control, Routing, and Congestion Control (JOPRC)
problem:
\begin{equation}\label{eq:DistFlowCtrlRTAndPowerCtrlProblem}
\begin{array}{lll}
\vspace{1mm} \textrm{minimize} \; &\displaystyle{\sum_{(i,j) \in
{\cal E}}D_{ij}(C_{ij},F_{ij}) + \sum_{w \in {\cal W}}
B_w(F_{wb})}
\\
\vspace{1mm} \textrm{subject to} &\phi_{ij}(w) \ge 0,~\phi_{wb}
\ge 0, \; &\forall w \in {\cal W} ~\textrm{and}~(i,j) \in {\cal E}
\\
\vspace{1mm} &\displaystyle{\sum_{j \in {\cal O}(i)}{\phi_{ij}(w)} = 1} &\forall w \in {\cal W} ~\textrm{and}~ i \ne
O(w),D(w)
\\
\vspace{1mm} &\displaystyle{\sum_{j \in {\cal O}(i)}{\phi_{ij}(w)}
+ \phi_{wb} = 1}, &\forall w \in {\cal W} ~\textrm{and}~ i = O(w)
\\
\vspace{1mm} &\eta_{ij} \ge 0, &\forall (i,j) \in {\cal E}
\\
\vspace{1mm} &\displaystyle{\sum_{j \in {\cal O}(i)}{\eta_{ij}} =
1}, &\forall i \in {\cal N}
\\
\vspace{1mm} & \gamma_{i} \le 1, &\forall i \in {\cal N}
\\
\end{array}
\end{equation}
where link flow rates and capacities are determined by the
optimization variables as
\begin{equation}\label{eq:FlowCapacityDependenceRedef}
\begin{array}{ll}
\vspace{1mm} F_{ij} = \displaystyle\sum_{w \in {\cal W}}{t_i(w)
\cdot \phi_{ij}(w)}, \; &\forall (i,j) \in {\cal E}
\\
\vspace{1mm} F_{wb} = t_{O(w)}(w) \cdot \phi_{wb}, &\forall w \in
{\cal W}
\\
\vspace{1mm} t_i(w) = \left\{
\begin{array}{ll}
                            \vspace{1mm} \bar r_w, \; &\textrm{if}~i =
                            O(w) \\
                            \displaystyle\sum_{j \in
                            {\cal I}(i)}{t_j(w) \cdot \phi_{ji}(w)},
                            &\textrm{if}~i \ne
                            O(w)
                            \end{array} \right. \\
\vspace{1mm} C_{ij} = C\left(\displaystyle\frac{G_{ij} (\bar P_i)^{\gamma_i} \eta_{ij}}{
G_{ij} (\bar P_i)^{\gamma_i} \displaystyle\sum_{k \ne j}{\eta_{ik}} + \sum_{m \ne
i}{G_{mj} (\bar
P_m)^{\gamma_m}} + N_j }\right), &\forall (i,j) \in {\cal E}. \\
\end{array}
\end{equation}

The optimality conditions for~\eqref{eq:DistFlowCtrlRTAndPowerCtrlProblem} are the same as in Theorem
\ref{thm:OptimalityCondition}, except that the optimal routing condition for all source nodes are modified. For all $w
\in {\cal W}$ and $i = O(w)$, these conditions are
\begin{equation}\label{eq:FlowCtrlOptCond}
\begin{array}{ll}
\vspace{1mm} \delta\phi_{ij}(w) = \lambda_i(w), \;&\textrm{if}~\phi_{ij}(w) > 0
\\
\vspace{1mm} \delta\phi_{ij}(w) \ge \lambda_i(w), &\textrm{if}~\phi_{ij}(w) = 0
\\
\vspace{1mm} \delta\phi_{wb} = \lambda_i(w),
&\textrm{if}~\phi_{wb}
> 0
\\
\vspace{1mm} \delta\phi_{wb} \ge \lambda_i(w),
&\textrm{if}~\phi_{wb} = 0
\\
\end{array}
\end{equation}
for some constant $\lambda_i(w)$, where the marginal cost $\delta\phi_{wb}$ of the overflow link is defined as
\begin{equation}\label{eq:OverflowMargCost}
\delta\phi_{wb} = B_w'(F_{wb}), \;\forall w \in {\cal W}.
\end{equation}

The proof of the above result is almost a repetition of the argument
for Theorem~\ref{thm:OptimalityCondition}, and is skipped here. This
optimality condition can be interpreted as follows: the flow of a
session is routed only onto minimum-marginal-cost path(s) and the
marginal cost of rejecting traffic is equal to the marginal cost of
the path(s) with positive flow.

The distributed algorithms for achieving the optimum are the same as in Section~\ref{sec:ConvergenceOfAlgorithms},
except for changes at the source nodes.  To mark the difference, we recast the modified routing algorithm as a joint
congestion control/routing ($CR$) algorithm at the source nodes. At every iteration, it has the same scaled gradient
projection form:
\[
\boldsymbol{\phi}_i^{k + 1}(w) = CR(\boldsymbol\phi_i^k(w)) = \left[\boldsymbol\phi_i^k(w) - (M_i^k(w))^{-1} \cdot
\delta\boldsymbol\phi_i^k(w)\right]_{M_i^k(w)}^+.
\]
Notice that the definitions for $\bs\phi_i(w)$ and $\delta\bs\phi_i(w)$ now become $\bs\phi_i(w) \triangleq (\phi_{wb},
(\phi_{ij}(w))_{j \in {\cal O}(i)})$ and $\delta\bs\phi_i(w) \triangleq (\delta\phi_{wb}, (\delta\phi_{ij}(w))_{j \in
{\cal O}(i)})$. Accordingly, the scaling matrix $M_i^k(w)$ is expanded by one in dimension. \vspace{0.1in}

Observe that with the introduction of the virtual overflow link, we naturally find an
initial loop-free routing configuration for the $CR$ algorithm: $\phi_{wb} = 1$ for all
$w \in {\cal W}$. That is, the traffic is fully blocked. This configuration can be set up
independently by the source nodes, and is preferable to other loop-free startup
configurations, since it does not cause any potential transient overload on any link
inside the network. Due to the fact that the $RT$ algorithm outputs a loop-free
configuration if the input routing graph is loop-free~\cite{paper:Gal77}, we can assert
that at all iterations, the $CR$ algorithm yields loop-free updates.  Next, we note that
$CR(\cdot)$ is fully supported by the marginal-cost-message exchange protocol introduced
after the algorithms in Section~\ref{subsec:RT}, since the only extra measure is
$\delta\phi_{wb}$, which is obtainable locally at the source node.

\section{Numerical Experiments}\label{sec:Simulation}

In this section, we present the results of numerical experiments
which point to the superior performance of the node-based routing,
power allocation, and power control algorithms presented in Sections
\ref{sec:ConvergenceOfAlgorithms}.  First, we compare our routing
algorithm with the Ad hoc On Demand Distance Vector (AODV)
algorithm~\cite{paper:PR99} both in static networks and in networks
with changing topology and session demands.  Next, we assess the
performance of the power control (PC) algorithm when the power
control messages are propagated only locally.  Finally, we test the
robustness of our algorithms to noise and delay in the marginal cost
message exchange process. For all experiments, we adopt $D_{ij} =
\frac{F_{ij}}{C_{ij}-F_{ij}}$ as the link cost function.}

\subsection{Comparison of AODV and BRT in Static Networks}

{We first compare the average network cost\footnote{Recall that the
network cost is the sum of costs on all links.} trajectories
generated by the AODV algorithm and the Basic Routing (BRT)
Algorithm~\eqref{eq:BasicRT}-\eqref{eq:BasicRTUpdate} under a static
network setting. We also compare the cost trajectories of AODV and
BRT when they are iterated jointly with the Basic Power Allocation
(BPA) and Power Control (PC) algorithms.
\begin{figure}[h]
  \begin{center}
  \includegraphics[width=10cm]{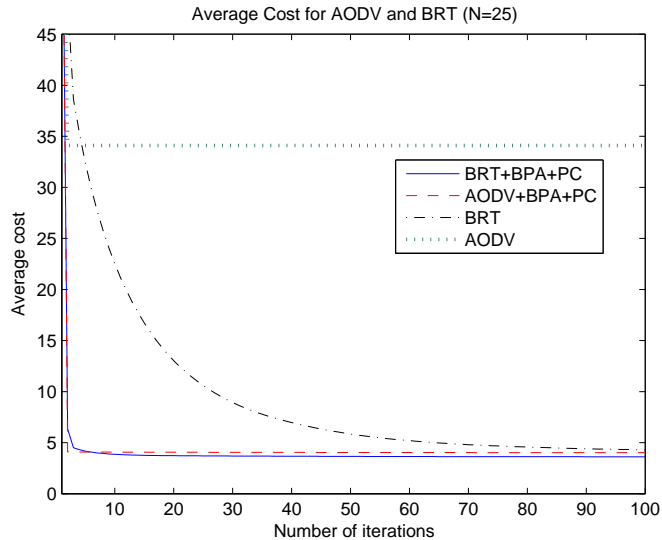}\\
  \caption{Average cost trajectories generated by AODV and BRT with and without BPA and PC.}\label{fig:AODVvsRT}
  \end{center}
\end{figure}
The trajectories in Figure~\ref{fig:AODVvsRT} are obtained from averaging 20 independent
simulations of the AODV, BRT, BPA and PC algorithms on the same network with the same
session demands. For each simulation, the network topology and the session demands are
randomly generated as follows.}

For a fixed number of nodes $N=25$, let the $N$ nodes be uniformly distributed in a disc
of unit radius. There exists a link between nodes $i$ and $j$ if their distance $d(i,j)$
is less than $0.5$. The path gain is modelled as $G_{ij} = d(i,j)^{-4}$. We use capacity
function $C_{ij} = \log(K \cdot SINR_{ij}) $, where $K$ represents the processing gain.
In our experiment, $K$ is taken to be $10^5$. All nodes are subject to a common power
constraint $P_i \le \bar P = 100$ and AWGN of power $N_i = 0.1$. Each node generates
traffic input to the network with probability $1/2$, and independently picks its
destination from the other $N-1$ nodes at random. In the experiments, we assume all
active sessions are inelastic, each with incoming rate determined independently according
to the uniform distribution on $[0,10]$.


{When the AODV and BRT algorithms are iterated without the BPA and
PC algorithms, we let every node transmit at the maximal power $\bar
P$ and evenly allocate the total power to its outgoing links. As we
can see from Figure~\ref{fig:AODVvsRT}, since AODV always seeks out
the minimum-hop paths for the sessions without consideration for the
network cost, convergence to its intended optimal routing takes only
a few iterations,\footnote{In all our simulations, one iteration
involves every node updating its routing, power allocation, and
power control variables once using the corresponding algorithms.}
while the BRT algorithm converges only asymptotically. However in
terms of network cost, BRT achieves the fundamental optimum and it
always outperforms AODV. The performance gap between the AODV and
BRT algorithms is significantly reduced by the introduction of the
BPA and PC algorithms.
In fact, the performance gains attributed to the BPA and PC
algorithms are so significant that using AODV along with BPA and PC
yields a total cost very close the optimal cost achievable by the
combination of BRT, BPA and PC.}

\subsection{Comparison of AODV and BRT with Changing Topology and
Session Demands}

\label{sec:Simulation_change}

{We next compare the performance of the AODV and the Basic Routing
Algorithm in a quasi-static network environment where network
conditions vary slowly relative to the time scale of algorithm
iterations. In particular, we study the effects of time-varying
topology and time-varying session demands.}

\begin{figure}[!h]
  \begin{center}
  \includegraphics[width=10cm]{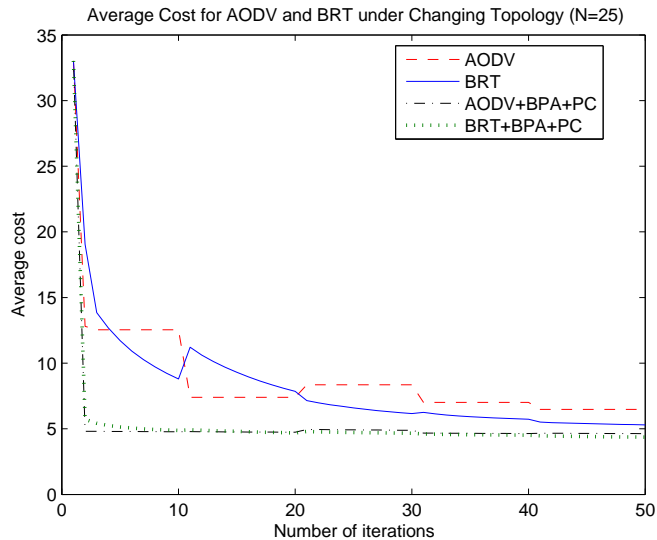}\\
  \caption{Average cost trajectories generated by AODV and BRT with and without BPA, PC under changing topology.}\label{fig:ChngTopology}
  \end{center}
\end{figure}

For each independent simulation, the network is initialized in the
same way as the previous experiment. After initialization, the
network topology changes after every 10 algorithm iterations. At
every changing instant, each node independently moves to a new
position selected according to a uniform distribution within a $0.1
\times 0.1$-square centered at the original location of that node.
We assume that the connectivity of the network remains
unchanged,\footnote{This is reasonable because nodes are assumed to
randomly move within their local area.} so that the movement of
nodes only causes variation in the channel gains $\{G_{ij}\}$.
{Figure~\ref{fig:ChngTopology} shows the average cost trajectories
generated by AODV and BRT with and without the power algorithms,
under the same topology changes.}
It can be seen from the figure that, relative to AODV, BRT adapts
very well to the time-varying topology. It is able to consistently
reduce the network cost after every topology change. In the long
run, BRT closes in on a routing that is almost optimal for all minor
topology changes produced by our movement model. In contrast, AODV
is not perceptive to the changes since it uses only hop counts as
the routing metric. As a result, the routing established by AODV is
never re-adjusted for the new topologies, and it yields higher cost
than the routing generated by BRT. {However, the performance of AODV
with BPA and PC is virtually as good as BRT with BPA and PC. Since
the power algorithms are highly adaptive to topology changes, they
almost completely make up the inability of AODV to adapt to topology
changes. }

\begin{figure}[!h]
  \begin{center}
  \includegraphics[width=10cm]{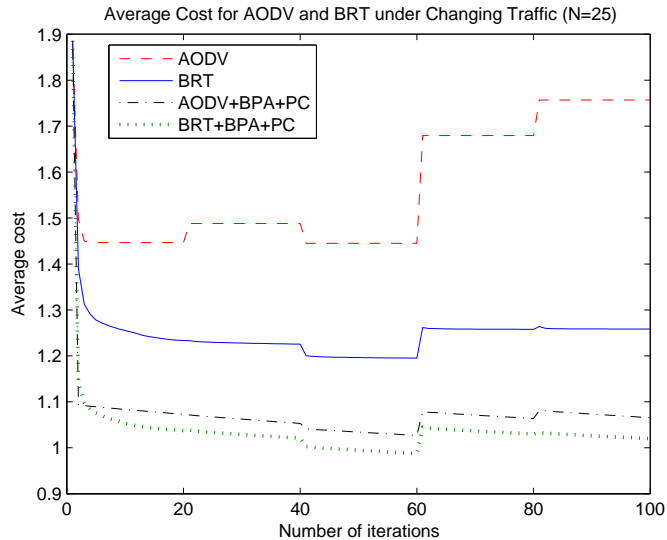}\\
  \caption{Average cost trajectories generated by AODV and BRT with and without BPA, PC under changing traffic demands.}\label{fig:ChngTraffic}
  \end{center}
\end{figure}

Figure~\ref{fig:ChngTraffic} compares the performance of AODV and BRT under time-varying
traffic demands.
{After the sessions are randomly initialized (in the same way as
above), we let the session rates fluctuate independently after every
10 iterations. At each instant of change, the new rate of a session
$w$ is determined by $\tilde r_w = \alpha_w r_w$ where the random
factor $\alpha_w$ is uniformly distributed from $0$ to $2$, and
$r_w$ is the original rate of $w$.  Again, BRT exhibits superior
adaptability compared to AODV.  BRT tends to establish a routing
almost optimal for all traffic demands generated by the above random
rate fluctuation model. On the other hand, the advantage of BRT over
AODV becomes less evident when they are implemented together with
the BPA and PC algorithms.}

\subsection{Power Control with Local Message Exchange}

{One major practical concern for the implementation of the Power
Control (PC) algorithm~\eqref{eq:GeneralPCUpdate} is that for every
iteration it requires each node to receive and process one message
from every other node in the network (cf.
Sec.~\ref{subsec:PCMsgEx}). As a result, the PC algorithm, when
exactly implemented, incurs communication overhead that scales
linearly with $N$. On the other hand, extensive simulations indicate
that the PC algorithm functions reasonably well even with message
exchange restricted to nearby nodes. One can understand this
phenomenon intuitively by inspecting the formula for the marginal
power control cost $\delta\gamma_i$~\eqref{eq:DeltaGamma}. Note that
the power control message from node $n$ is multiplied by $G_{in}$ on
the RHS~\eqref{eq:DeltaGamma}. Thus, for $n$ far from $i$, the
contribution of $MSG(n)$ to $\delta\gamma_i$ is negligible due to
the small factor $G_{in}$.}

{In the present experiment, The network and sessions are generated
randomly in the same way as before.  The routing is fixed according
to a minimum-hop criterion, and all nodes uniformly allocate power
on its outgoing links. We implement different approximate versions
of the PC algorithm where the power control messages are propagated
only locally. Each version of PC calculates the marginal power
control costs $\delta\gamma_i$ approximately by using power control
messages from a certain number of neighbors of $i$. To be specific,
the exact formula~\eqref{eq:DeltaGamma} is now approximated by \[
\frac{\delta\gamma_i}{P_i} \approx \sum_{j \in {\cal N}(i)} G_{ij}
MSG(j) + \sum_{n \in {\cal O}(i)} \delta\eta_{in} \cdot \eta_{in},
\]
where ${\cal N}(i)$ is the subset of nodes that are closest to $i$. The size of ${\cal
N}(i)$ varies from $1$ to $8$ for different versions of PC simulated in this experiment.
The network and sessions are generated randomly in the same as before.
Figure~\ref{fig:LocalPC} shows the cost trajectories obtained from averaging a number of
independent simulations.
\begin{figure}[!h]
  \begin{center}
  \includegraphics[width=12cm]{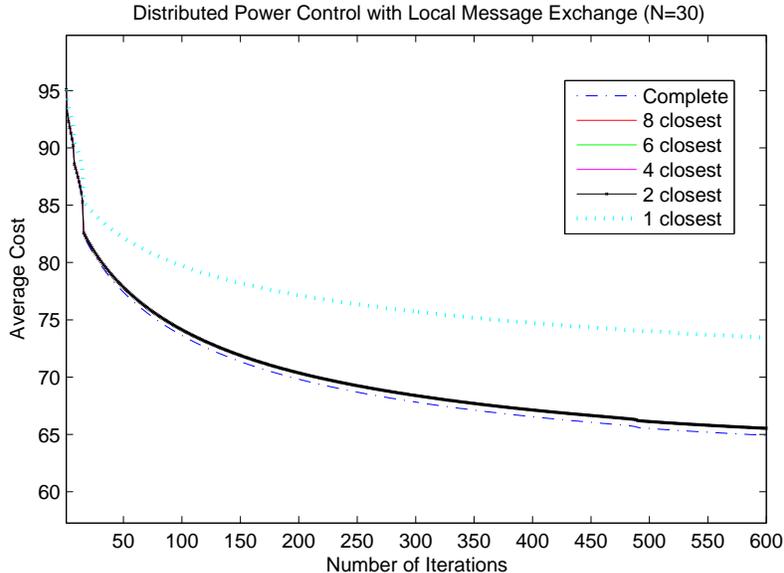}\\
  \caption{Average cost trajectories generated by PC with different message exchange scopes.}\label{fig:LocalPC}
  \end{center}
\end{figure}
For example, the dotted line represents the cost trajectory generated by the PC algorithm
that approximates the marginal cost $\delta\gamma_i$ using $MSG(j)$ only from the node
nearest to $i$. Results from Figure~\ref{fig:LocalPC} indicate that as long as the
computation of $\delta\gamma_i$ incorporates messages from at least two nearest
neighbors, the performance of PC is almost indistinguishable from that of PC with
complete message exchange.}

\subsection{Algorithms with Delayed and Noisy Messages}

{Finally, we simulate the joint application of the routing, power allocation, and power
control algorithms in the presence of delay and noise in the exchange of marginal cost
messages. We model the delay resulting from infrequent updates by the nodes.
Specifically, we let each node $i$ update routing message $\frac{\partial D}{\partial
r_i(w)}$ using~\eqref{eq:NodeMarginalRoutingCost2} only when it iterates
$RT(\bs\phi_i(w))$, and we let node $i$ update power control message $MSG(i)$
using~\eqref{eq:MSG} only when it iterates $PC(\gamma_i)$. As a consequence, the marginal
costs $\delta\phi_{ij}(w)$ and $\gamma_i$ have to be computed based on outdated
information from other nodes, as that information was last updated when the other nodes
last iterated.

In addition to delay, we assume messages are subject to noise such
that the message received is a random factor times the true
value.\footnote{The multiplicative noise is attributed to, for
instance, errors in estimating the state of the fading channel over
which marginal cost messages are sent.} Each message transmission is
subject to an independent random factor drawn from a uniform
distribution on $[1-\textrm{NoiseScale}, 1 + \textrm{NoiseScale}]$
where the parameter NoiseScale is taken to be $0.9$ in the
simulations shown in Figure~\ref{fig:DelayNoise}.
\begin{figure}[!h]
  \begin{center}
  \includegraphics[width=12cm]{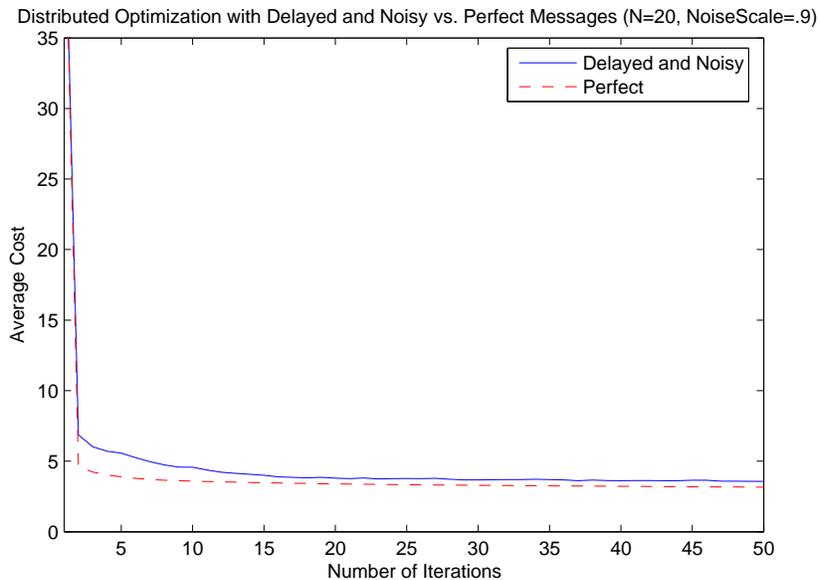}\\
  \caption{Average cost trajectories generated by BRT, BPA and PC with delayed and noisy vs. perfect messages.}\label{fig:DelayNoise}
  \end{center}
\end{figure}
Compared to using constantly updated and noiseless messages, the
algorithms with delayed and noisy message exchange converge to a
limit only slightly worse than the true optimum. }

\vspace{3mm}  In conclusion, the simulation results confirm that the
BRT, BPA and PC algorithms have fast and guaranteed convergence.
Moreover, they exhibit satisfactory convergence behavior under
changing network topology and traffic demands, as well as in the
presence of delay and noise in the marginal cost exchange process.
In particular, the PC algorithm performs reasonably well when power
control messages are propagated only locally.   All these results
attest to the practical applicability of our algorithms to real
wireless networks.

Finally, we note that the power allocation and power control
accounted for most of the cost reduction when the performance of RT
with BPA and PC was compared to that of AODV with BPA and PC.  This
points to the importance of jointly optimizing power control and
routing, and suggests that implementing the power allocation and
power control algorithms jointly with existing routing algorithms
can result in large performance gains.

\vspace{3mm}\section{\label{sec:Conclusion}Conclusion}\vspace{3mm}

We have presented a general flow-based analytical framework in which
power control, rate allocation, routing, and congestion control can
be jointly optimized to balance aggregate user utility and total
network cost in wireless networks.  A complete set of distributed
node-based scaled gradient projection algorithms are developed for
interference-limited networks where routing, power allocation, and
power control variables are iteratively adjusted at individual
nodes. We have explicitly characterized the appropriate scaling
matrices under which the distributed algorithms converge to the
global optimum from any initial point with finite cost.  It is shown
that the computation of these scaling matrices require only a
limited number of control message exchanges in the network.
Moreover, convergence does not depend on any particular ordering and
synchronization in implementing the algorithms at different nodes.

To enlarge the space of feasible controls, we relaxed the high-SINR
assumption for SINR-dependent link models by using the precise
capacity function for the problem of jointly optimizing routing and
power allocation.  We further extended the analytical framework to
consider wireless networks with general convex capacity region and
strictly quasiconvex link costs. It is proved that in this general
setting, an operating point satisfying equilibrium conditions is
Pareto optimal. Next, we showed that congestion control can be
seamlessly incorporated into our framework, in the sense that the
problem of jointly optimal power control, routing, and congestion
control can be made equivalent to a problem involving power control
and routing in a virtual wireless network with the addition of
virtual overflow links. Finally, results from numerical experiments
indicate that the distributed network algorithms have superior
performance relative to existing schemes, that the algorithms have
good adaptability to time-varying network conditions, and that they
are robust to delay and noise in the control message exchange
process.

\newpage
\appendix

\subsection{Proof of Lemma~\ref{lma:LinkNodeRateRelation}}\label{app:LinkNodeRateRelation}

Multiplying both sides of \eqref{eq:NodeMarginalRoutingCost2} for $i = O(w)$ by $r_w$ and summing over all $w \in {\cal
W}$, we have
\begin{eqnarray*}
\displaystyle{\sum_{w \in {\cal W}}\frac{\partial D}{\partial r_{O(w)}(w)} \cdot r_w} &=& \displaystyle{\sum_{w \in
{\cal W}}\sum\limits_{k \in {\cal O}(O(w))} r_w \phi_{O(w)k}(w)\left[\frac{\partial D_{O(w)k}}{\partial
F_{O(w)k}}(C_{O(w)k},F_{O(w)k}) + \frac{\partial D}{\partial r_k(w)}\right]}
\\
&=& \displaystyle{\sum_{w \in {\cal W}} \sum_{k \in {\cal O}(O(w))} f_{O(w)k}(w) \frac{\partial D_{O(w)k}}{\partial
F_{O(w)k}}(C_{O(w)k},F_{O(w)k}) }
\\
&&+ \displaystyle{\sum_{w \in {\cal W}} \sum_{k \in {\cal O}(O(w))} \sum_{j \in {\cal O}(k)} f_{O(w)k}(w)
\phi_{kj}(w)\left[\frac{\partial D_{kj}}{\partial F_{kj}}(C_{kj},F_{kj}) + \frac{\partial D}{\partial r_j(w)}\right]}.
\end{eqnarray*}
Expand the term $\frac{\partial D}{\partial r_j(w)}$ repeatedly until $j = D(w)$, where $\frac{\partial D}{\partial
r_j(w)} = 0$. Then, use the flow conservation relation $t_k(w) = \sum_{i \in {\cal I}(k)} f_{ik}(w)$ for $k \ne O(w)$
to successively factor out terms $t_k(w) \phi_{kj}(w) = f_{kj}(w)$. Finally, noticing that the outermost summation
yields
\[
F_{ik} = \sum_{w \in {\cal W}}{f_{ik}(w)},
\]
we obtain the equality of the LHS and RHS of \eqref{eq:LinkNodeRateRelation}. \qed

\subsection{Proof of Lemma~\ref{lma:PhiHessianUppBnd}}\label{app:PhiHessian}

For simplicity, we suppress session index $w$ and iteration index $k$. For $i \ne D(w)$, the entries of
$H_{\boldsymbol\phi_i}^\lambda$ corresponding to subspace $\left\{\boldsymbol v_i : \sum_{j \in {\cal AN}_i}{v_{ij}} =
0 \right\}$ are as follows. For $k,j \in {\cal AN}_i$,
\begin{equation}\label{eq:RTHessianEntries}
\begin{array}{l}
\vspace{1mm} \displaystyle{\left[H_{\boldsymbol\phi_i}^\lambda\right]_{kk} = \frac{\partial^2 D}{\partial \phi_{ik}^2}
= t_i^2 \left[\frac{\partial^2 D_{ik}}{\partial F_{ik}^2}  + \frac{\partial^2 D}{\partial r_k^2}\right]},
\\
\vspace{1mm} \displaystyle{\left[H_{\boldsymbol\phi_i}^\lambda\right]_{kj} = \frac{\partial^2 D}{\partial \phi_{ik}
\partial \phi_{ij}} = t_i^2 \frac{\partial^2 D}{\partial r_k \partial r_j }}, \; k \ne j.
\end{array}
\end{equation}
Note that the terms $\frac{\partial^2 D_{ik}}{\partial F_{ik}^2}$ are locally measurable. Thus, in the following, we
deal only with the terms $\frac{\partial^2 D}{\partial r_k^2}$ and $\frac{\partial^2 D}{\partial r_k
\partial r_j }$ for $k,j \in {\cal AN}_i$. In \cite{paper:BGG84}, the authors provide the
following useful expression:
\begin{equation}\label{eq:RTSecDerivExpression}
\frac{\partial^2 D}{\partial r_k \partial r_j} = \sum_{(m,n) \in {\cal E}}{q_{mn}(k) q_{mn}(j)\frac{\partial^2
D_{mn}}{\partial F_{mn}^2}},
\end{equation}
where $q_{mn}(k)$ denotes the fraction of a unit flow originating at node $k$ that goes through link $(m,n)$. By the
Cauchy-Schwarz Inequality,
\begin{equation}\label{eq:RTSecDerivCauchyIneq}
\frac{\partial^2 D}{\partial r_k \partial r_j} \le \sqrt {\frac{\partial^2 D}{\partial r_k^2} \frac{\partial^2
D}{\partial r_j^2}}.
\end{equation}
Multiplying $H_{\boldsymbol\phi_i}^\lambda$ on the left and right with non-zero vector $\boldsymbol v_i$, we have
\begin{equation}\label{eq:RTHessianUpperBound}
\begin{array}{ll}
\vspace{1mm} \boldsymbol v_i' \cdot H^\lambda_{\boldsymbol\phi_i} \cdot \boldsymbol v_i &\displaystyle{= t_i^2
\left[\sum_{j \in {\cal AN}_i}{\left(\frac{\partial^2 D_{ij}}{\partial F_{ij}^2} + \frac{\partial^2 D}{\partial
r_j^2}\right) v_{ij}^2 } + \sum_{\scriptstyle j,k \in {\cal AN}_i \hfill \atop
  \scriptstyle{\textrm{and}}~j \ne k \hfill}{\frac{\partial^2 D}{\partial r_j \partial r_k} v_{ij} v_{ik} }\right]}
\\
\vspace{1mm} &\displaystyle{\le t_i^2 \left[\sum_{j \in {\cal AN}_i}{\left(\frac{\partial^2 D_{ij}}{\partial F_{ij}^2}
+ \frac{\partial^2 D}{\partial r_j^2}\right) v_{ij}^2 } + \sum_{\scriptstyle j,k \in {\cal AN}_i \hfill \atop
  \scriptstyle{\textrm{and}}~j \ne k \hfill}{\sqrt{\frac{\partial^2 D}{\partial r_j^2}\frac{\partial^2 D}{\partial r_k^2}} |v_{ij}| |v_{ik}|}\right]}
\\
\vspace{1mm} &\displaystyle{= t_i^2 \left[\sum_{j \in {\cal AN}_i}{\frac{\partial^2 D_{ij}}{\partial F_{ij}^2}
v_{ij}^2} + \left(\sum_{j \in {\cal AN}(i)}{\sqrt{\frac{\partial^2 D}{\partial r_j^2}} |v_{ij}|}\right)^2\right]}
\\
\vspace{1mm} &\displaystyle{\le t_i^2 \left[\sum_{j \in {\cal AN}_i}{\left(\frac{\partial^2 D_{ij}}{\partial F_{ij}^2}
+ |{\cal AN}_i| \frac{\partial^2 D}{\partial r_j^2}\right) v_{ij}^2 }\right]}
\\
\vspace{1mm} &= t_i^2 \cdot \boldsymbol v_i' \cdot \tilde H_{\boldsymbol\phi_i } \cdot \boldsymbol v_i,
\end{array}
\end{equation}
where the last inequality follows from applying the Cauchy-Schwarz Inequality to the inner product of vector
$(\sqrt{\frac{\partial^2 D}{\partial r_j^2}} |v_{ij}|)_{j \in {\cal AN}_i}$ with the all-one vector of the same
dimension. Thus, the Hessian matrix is upper bounded by the positive definite matrix $t_i^2 \cdot \tilde
H_{\boldsymbol\phi_i}$, where
\[ \tilde H_{\boldsymbol\phi _i} = {\rm diag}\left\{\frac{\partial^2 D_{ij}}{\partial F_{ij}^2} + |{\cal AN}_i| \frac{\partial^2 D}{\partial r_j^2}\right\}_{j \in {\cal
AN}_i}.
\]
Note that $|{\cal AN}_i|$ is the non-blocked out-degree of node $i$ at the current iteration. Also, note that
$\frac{\partial^2 D_{ij}}{\partial F_{ij}^2}$ and $\frac{\partial^2 D}{\partial r_j^2}$ are evaluated at
$\lambda\boldsymbol\phi_i^k + (1-\lambda)\boldsymbol\phi_i^{k+1} \triangleq \boldsymbol\phi_i^{k,\lambda}$ for some
$\lambda \in [0,1]$.

There are various ways of producing approximate upper bounds on the
diagonal terms in $\tilde H_{\boldsymbol\phi_i}$. Bertsekas et
al.\cite{paper:BGG84} propose a message propagation scheme in the
network where the propagated messages are upper bounds of
$\frac{\partial^2 D}{\partial r_j^2}$ calculated by the
corresponding nodes using the bounds provided by their downstream
nodes. Such information exchange can be implemented in the same
manner as that used for propagating the first derivative cost
information $\frac{\partial D}{\partial r_j(w)}$.

To limit communication and computational complexity in wireless networks, we present another scheme which requires less
overhead but nevertheless yields reasonable bounds. Let $A \triangleq \max_{(m,n) \in {\cal E}}\left.\frac{\partial^2
D_{mn}}{\partial F_{mn}^2}\right|_{F_{mn}=F_{mn}(\boldsymbol\phi_i^{k,\lambda})}$. We have
\[
\frac{\partial^2 D}{\partial r_j^2} = \sum_{(m,n)} q_{mn}(j)^2 \left.\frac{\partial^2 D_{mn}}{\partial
F_{mn}^2}\right|_{F_{mn}=F_{mn}(\boldsymbol\phi_i^{k,\lambda})} \le A \sum_{(m,n)} q_{mn}(j)^2 \le A \sum_{(m,n)}
q_{mn}(j) \le A \cdot h_j.
\]
To prove the last inequality, we focus on paths connecting $j$ and
$D(w)$ induced by any loop-free routing pattern. Denote the
collection of such paths by ${\cal P}_j$ and let $\delta_{p}$ be the
increase of path flow on $p \in {\cal P}_j$ as the result of a unit
increment of input at node $j$. Thus, we have $\sum_{p \in {\cal
P}_j} \delta_{p} = 1$ and
\[
q_{mn}(j) = \sum_{p \in {\cal P}_j: (m,n) \in p} \delta_{p}.
\]
Summing over all $(m,n)$, we obtain
\[
\sum_{(m,n)} q_{mn}(j) = \sum_{(m,n)} \sum_{p \in {\cal P}_j: (m,n) \in p} \delta_{p} = \sum_{p \in {\cal P}_j}
\sum_{(m,n) \in p} \delta_{p} = \sum_{p \in {\cal P}_j} h_j(p) \cdot \delta_{p} \le h_j,
\]
where $h_j(p)$ is the number of hops on path $p$ and $h_j = \max_{p \in {\cal P}_j} h_j(p)$.


Now since the initial network cost is no greater than $D^0$ and the cost is strictly reduced with every iteration of
$RT(\cdot)$ until conditions \eqref{eq:RoutingOptCond1}-\eqref{eq:RoutingOptCond2} are satisfied, for two consecutive
steps $k$ and $k+1$, on any link $(m,n)$, $\left.D_{mn}(C_{mn},F_{mn})\right|_{F_{mn}=F_{mn}(\boldsymbol\phi_i^{k})}
\le D^0$ and $\left.D_{mn}(C_{mn},F_{mn})\right|_{F_{mn}=F_{mn}(\boldsymbol\phi_i^{k+1})} \le D^0$. By convexity of the
cost function, $\left.D_{mn}(C_{mn},F_{mn})\right|_{F_{mn}=F_{mn}(\boldsymbol\phi_i^{k,\lambda})} \le D^0$, for all
$\lambda \in [0,1]$. Therefore by the definitions of $A(D^0)$ and $A$, we have $A \le A(D^0)$. Also
$\left.\frac{\partial^2 D_{ij}}{\partial F_{ij}^2}\right|_{F_{ij}=F_{ij}(\boldsymbol\phi_i^{k,\lambda})} \le
A_{ij}(D^0)$ for the same reason. Putting all these results together, we can further upper bound $t_i^2 \cdot \tilde
H_{\boldsymbol\phi_i}$ by \[{t_i}^2{\rm diag}\left\{\left( A_{ij}(D^0) + |{\cal AN}_i| h_j A(D_0)\right)_{j \in {\cal
AN}_i}\right\}.\] Thus, the proof is complete. \qed

\subsection{Proof of Lemma~\ref{lma:EtaHessianUppBnd}}\label{app:EtaHessian}

Constrained in ${\cal V}_{\bs\eta_i}$,  the transmission powers,
hence the capacities, of node $i$'s outgoing links are subject to
change. We can therefore focus on $C_{ij}$ for $j \in {\cal O}(i)$
and take it as a function of $\eta_{ij}$:
\[
C_{ij} = C(SINR_{ij}) = C\left(\frac{ G_{ij} P_i \eta_{ij}}{G_{ij} P_{i} (1 - \eta_{ij}) + \sum_{m \ne i} G_{mj} P_{m}
+ N_j}\right) \triangleq C_{ij}(\eta_{ij}).
\]
It can be seen that the Hessian matrix $H_{\boldsymbol\eta_i}^{k, \lambda}$ is diagonal. Omitting the superscript $(k,
\lambda)$, we can write the diagonal terms as \begin{small} \beas \left[H_{\boldsymbol\eta_i}\right]_{jj} &=&
 \frac{\partial^2 D_{ij}}{\partial C_{ij}^2} \left[ C'(x_{ij}) \frac{G_{ij}
P_i}{IN_{ij}} (1 + x_{ij})\right]^2 + \frac{\partial D_{ij}}{\partial C_{ij}} \left[ C''(x_{ij}) \left( \frac{ G_{ij}
P_i}{IN_{ij}} (1 + x_{ij}) \right)^2 + C'(x_{ij}) \frac{G_{ij}^2 P_{i}^2}{IN_{ij}^2} (2 + 2 x_{ij}) \right]
\\
&=&  \left(\frac{P_i}{P_{ij}}\right)^2 \left\{ \frac{\partial^2 D_{ij}}{\partial C_{ij}^2} C'(x_{ij})^2 x_{ij}^2 (1 +
x_{ij})^2  + \frac{\partial D_{ij}}{\partial C_{ij}} \left[ C''(x_{ij}) x_{ij}^2 (1 + x_{ij})^2 + 2 C'(x_{ij}) x_{ij}^2
(1 + x_{ij})  \right] \right\}. \eeas \end{small} Because $\eta_{ij}$ must be lower bounded by $\underline\eta_{ij}>0$
for which  $D_{ij}(C_{ij}(\underline\eta_{ij}), F_{ij}^k) = D_i^k$, we have $\eta_{ij}^{k, \lambda} >
\underline\eta_{ij}$ and $x_{ij}^{k, \lambda}$ is bounded as
\[
x_{ij}^{min} = \frac{ G_{ij} P_i \underline\eta_{ij}} {G_{ij} P_{i} (1 - \underline\eta_{ij}) + \sum_{m \ne i} G_{mj}
P_{m} + N_j} \le x_{ij}^{k, \lambda} \le \frac{G_{ij} P_i} {\sum_{m \ne i} G_{mj} P_{m} + N_j} = x_{ij}^{max},
\]
Thus, by dropping the negative term $2 \frac{\partial D_{ij}}{\partial C_{ij}} C'(x_{ij}) x_{ij}^2 (1 + x_{ij}) $ and
recalling the definition of $\beta_{ij}$ in~\eqref{eq:beta}, the diagonal term can be bounded as \begin{small} \beas
\left[H_{\boldsymbol\eta_i}\right]_{jj} &\le& \left( \frac{1}{\eta_{ij}^{k,\lambda}} \right)^2 \left[ \frac{\partial^2
D_{ij}}{\partial C_{ij}^2} \max_{x_{ij}^{min} \le x \le x_{ij}^{max}} \{ C'(x)^2 x^2 (1 + x)^2 \} + \frac{\partial
D_{ij}}{\partial C_{ij}} \min_{x_{ij}^{min} \le x \le
x_{ij}^{max}} \{ C''(x) x^2 (1 + x)^2 \} \right] \\
&<& \beta_{ij}. \eeas \end{small} Therefore, the lemma follows. \qed

\subsection{Proof of Lemma~\ref{lma:GammaHessianUppBnd}}\label{app:GammaHessian}

For brevity, we suppress the superscript $(k,\lambda)$ wherever it arises.  The Hessian matrix under consideration has
diagonal terms
\begin{small} \beas  [H_{\bs\gamma}]_{ii} &=& \frac{\partial^2 D}{\partial \gamma_i^2} \\
&=&  \bar S_i^2 \left\{ \sum_{m \ne i} \sum_{j \in {\cal O}(m)} \frac{\partial^2 D_{mj}}{\partial C_{mj}^2} (C'_{mj})^2
x_{mj}^2 \left(\frac{G_{ij} P_i}{IN_{mj}}\right)^2 + \frac{\partial D_{mj}}{\partial C_{mj}}
 (C_{mj}'' x_{mj}^2 + 2 C_{mj}' x_{mj}) \left(\frac{G_{ij} P_i}{IN_{mj}}\right)^2 \right. \\
 && - \frac{\partial D_{mj}}{\partial C_{mj}} C_{mj}' x_{mj} \frac{G_{ij}
P_i}{IN_{mj}} + \sum_{j \in {\cal O}(i)} \frac{\partial^2 D_{ij}}{\partial C_{ij}^2} C_{ij}'^2 x_{ij}^2
\left(1 - \frac{ G_{ij} (1-\eta_{ij}) P_i}{IN_{ij}}\right)^2 \\
&&  + \frac{\partial D_{ij}}{\partial C_{ij}} \left[ (C_{ij}'' x_{ij}^2 + C_{ij}' x_{ij}) \left(1 - \frac{ G_{ij}
(1-\eta_{ij}) P_i}{IN_{ij}} \right)^2 + C_{ij}' x_{ij} \left( \frac{ G_{ij} (1-\eta_{ij})
P_i}{IN_{ij}} \right)^2 \right. \\
&& \left. \left. - C_{ij}' x_{ij} \frac{ G_{ij} (1-\eta_{ij}) P_i}{IN_{ij}} \right] \right\} \\
&<& \bar S_i^2 \left\{ \sum_{m \ne i} \sum_{j \in {\cal O}(m)} \frac{\partial^2 D_{mj}}{\partial C_{mj}^2} (C'_{mj})^2
x_{mj}^2 \left(\frac{G_{ij} P_i}{IN_{mj}}\right)^2 + \frac{\partial D_{mj}}{\partial C_{mj}}
 (C_{mj}'' x_{mj}^2 + 2 C_{mj}' x_{mj}) \left(\frac{G_{ij} P_i}{IN_{mj}}\right)^2 \right. \\
 &&  + \sum_{j \in {\cal O}(i)} \frac{\partial^2 D_{ij}}{\partial C_{ij}^2} C_{ij}'^2 x_{ij}^2
\left(1 - \frac{ G_{ij} (1-\eta_{ij}) P_i}{IN_{ij}}\right)^2 \\
&&  + \frac{\partial D_{ij}}{\partial C_{ij}} \left[ (C_{ij}'' x_{ij}^2 + C_{ij}' x_{ij}) \left(1 - \frac{ G_{ij}
(1-\eta_{ij}) P_i}{IN_{ij}} \right)^2 + C_{ij}' x_{ij} \left( \frac{ G_{ij} (1-\eta_{ij})
P_i}{IN_{ij}} \right)^2 \right] \\
&& \left. - \sum_{(m,j) \in {\cal E}} \frac{\partial D_{mj}}{\partial C_{mj}} C_{mj}' x_{mj}  \right\}. \eeas
\end{small} The off-diagonal terms of the Hessian are \begin{small} \beas  \left[H_{\boldsymbol\gamma}\right]_{il} &=&
\frac{\partial^2 D}{\partial \gamma_i
\partial \gamma_l}
\\
&=&  \bar S_i \bar S_l \left\{\sum_{j \in {\cal O}(i)} \frac{\partial^2 D_{ij}}{\partial C_{ij}^2} (C'_{ij})^2
x_{ij}^2 \left(1 - \frac{ G_{ij} (1-\eta_{ij}) P_i}{IN_{ij}}\right) \frac{-G_{lj} P_l}{IN_{ij}} \right. \\
&& + \frac{\partial D_{ij}}{\partial C_{ij}} \left[ (C_{ij}'' x_{ij}^2 +  C_{ij}' x_{ij}) \left(1 - \frac{ G_{ij}
(1-\eta_{ij}) P_i}{IN_{ij}} \right) \frac{-G_{lj} P_l}{IN_{ij}} + (C_{ij}' x_{ij}) \frac{ G_{ij} G_{lj} (1-\eta_{ij})
P_i P_l}{IN_{ij}^2}\right]
\\
&&  + \sum_{j \in {\cal O}(l)} \frac{\partial^2 D_{lj}}{\partial C_{lj}^2} (C'_{lj})^2 x_{lj}^2 \left(1
- \frac{ G_{lj} (1-\eta_{lj}) P_l}{IN_{lj}}\right) \frac{-G_{ij} P_i}{IN_{lj}} \\
&& + \frac{\partial D_{lj}}{\partial C_{lj}} \left[ (C_{lj}'' x_{lj}^2 +  C_{lj}' x_{lj}) \left(1 - \frac{ G_{lj}
(1-\eta_{lj}) P_l}{IN_{lj}} \right) \frac{-G_{ij} P_i}{IN_{lj}} + (C_{lj}' x_{lj}) \frac{ G_{lj} G_{ij} (1-\eta_{lj})
P_l P_i}{IN_{lj}^2}\right]
\\
&& \left. + \sum_{m \ne i,l} {\sum_{j \in {\cal O}(m)} \frac{\partial^2 D_{mj}}{\partial C_{mj}^2} (C'_{mj})^2 x_{mj}^2
\frac{G_{ij} P_i G_{lj} P_l}{IN_{mj}^2} + \frac{\partial D_{mj}}{\partial C_{mj}} (C_{mj}'' x_{mj}^2 + 2 C_{mj}'
x_{mj}) \frac{G_{ij} P_i G_{lj} P_l}{IN_{mj}^2}} \right\}. \eeas \end{small}

Construct the following vectors for each $i \in {\cal N}$.
\[
U_i \triangleq \left\{ \begin{array}{ll} \vspace{1mm} [U_i]_{(m,j)} = \displaystyle{\sqrt{\frac{\partial^2
D_{mj}}{\partial C_{mj}^2}} C_{mj}' x_{mj} \frac{G_{ij} P_i }{IN_{mj}} \bar S_i v_i}, \; &\forall m \ne
i~\textrm{and}~j \in {\cal O}(m)
\\
\vspace{1mm} [U_i]_{(i,j)} = \displaystyle{-\sqrt{\frac{\partial^2 D_{ij}}{\partial C_{ij}^2}} C_{ij}' x_{ij} \left(1 -
\frac{ G_{ij} (1-\eta_{ij}) P_i}{IN_{ij}}\right) \bar S_i v_i}, &\forall j \in {\cal O}(i)
\end{array} \right.
\]
and
\[
T_i \triangleq \left\{ \begin{array}{ll} \vspace{1mm} [T_i]_{(m,j)} = \displaystyle{\sqrt{-\frac{\partial
D_{mj}}{\partial C_{mj}} C_{mj}' x_{mj} } \frac{G_{ij} P_i }{IN_{mj}} \bar S_i v_i }, \; &\forall m \ne
i~\textrm{and}~j \in {\cal O}(m)
\\
\vspace{1mm} [T_i]_{(i,j)} = \displaystyle{\sqrt{-\frac{\partial D_{ij}}{\partial C_{ij}} C_{ij}' x_{ij} } \frac{G_{ij}
(1-\eta_{ij}) P_i }{IN_{ij}} \bar S_i v_i }, &\forall j \in {\cal O}(i)
\end{array} \right. .
\]
\[
W_i \triangleq \left\{ \begin{array}{ll} \vspace{1mm} [W_i]_{(m,j)} = \displaystyle{\sqrt{ \frac{\partial
D_{mj}}{\partial C_{mj}} (C''_{mj} x_{mj}^2 + C_{mj}' x_{mj}) } \frac{G_{ij} P_i }{IN_{mj}} \bar S_i v_i }, \; &\forall
m \ne i~\textrm{and}~j \in {\cal O}(m)
\\
\vspace{1mm} [W_i]_{(i,j)} = \displaystyle{ - \sqrt{ \frac{\partial D_{ij}}{\partial C_{ij}} (C''_{ij} x_{ij}^2 +
C_{ij}' x_{ij}) } \left( 1 - \frac{G_{ij}  (1-\eta_{ij}) P_i }{IN_{ij}} \right) \bar S_i v_i }, &\forall j \in {\cal
O}(i)
\end{array} \right. .
\]
Thus, for any nonzero vector $\bs v \in \mathbb{R}^{|{\cal N}|}$ we have \beas && \bs v' \cdot H_{\bs\gamma}
\cdot \bs v \\
&<& \sum_{i,l \in {\cal N}}{U_i' \cdot U_l } + \sum_{i,l \in {\cal N}}{W_i' \cdot W_l } - \sum_{i,l \in {\cal N}}{T_i'
\cdot T_l} + \sum_{i \in {\cal N}}{\sum_{(m,j) \in {\cal E}}{-\frac{\partial
D_{mj}}{\partial C_{mj}} C_{mj}' x_{mj} \bar S_i^2 v_i^2 }}\\
&\le& \sum_{i,l \in {\cal N}}{\|U_i\| \cdot \|U_l\|} + \sum_{i,l \in {\cal N}}{\|W_i\| \cdot \|W_l\|} + \sum_{i \in
{\cal N}}{\sum_{(m,j) \in {\cal E}}{-\frac{\partial D_{mj}}{\partial C_{mj}} C_{mj}' x_{mj} \bar
S_i^2 v_i^2 }} \\
&\le & \frac{1}{2} \sum_{i,l \in {\cal N}}{\left(\|U_i\|^2 + \|U_l\|^2 + \|W_i\|^2 + \|W_l\|^2 \right)}  + \sum_{i \in
{\cal N}}{\sum_{(m,j) \in {\cal E}}{-\frac{\partial D_{mj}}{\partial C_{mj}} C_{mj}' x_{mj} \bar
S_i^2 v_i^2 }} \\
&= & |{\cal N}| \sum_{i \in {\cal N}} \left( \|U_i\|^2 + \|W_i\|^2 \right) + \sum_{i \in {\cal
N}}{\sum_{(m,j) \in {\cal E}}{-\frac{\partial D_{mj}}{\partial C_{mj}} C_{mj}' x_{mj} \bar S_i^2 v_i^2 }} \\
&= & \sum_{i \in {\cal N}} v_i^2 \bar S_i^2 \left\{|{\cal N}| \left[ \sum_{j \in {\cal O}(i)} \left( \frac{\partial^2
D_{ij}}{\partial C_{ij}^2} (C'_{ij})^2 x_{ij}^2 + \frac{\partial D_{ij}}{\partial C_{ij}} (C''_{ij} x_{ij}^2 + C'_{ij}
x_{ij}) \right) \left(1 - \frac{ G_{ij} (1-\eta_{ij}) P_i}{IN_{ij}} \right)^2
\right. \right. \\
&& \left. + \sum_{m \ne i} {\sum_{j \in {\cal O}_m} \left( \frac{\partial^2 D_{mj}}{\partial C_{mj}^2} (C'_{mj})^2
x_{mj}^2 + \frac{\partial D_{mj}}{\partial C_{mj}} (C''_{mj} x_{mj}^2 + C'_{mj} x_{mj}) \right) \left(\frac{G_{ij}
P_i}{IN_{mj}}\right)^2} \right]
\\
&{}&+   \left.\sum_{(m,j) \in {\cal E}} -\frac{\partial D_{mj}}{\partial C_{mj}} C'_{mj} x_{mj} \right\}. \eeas

To get a simpler upper bound, notice that $\left(1 - \frac{ G_{ij} (1-\eta_{ij}) P_i}{IN_{ij}} \right)^2 \le 1$,
$\left(\frac{G_{ij} P_i}{IN_{mj}}\right)^2 < 1$, and by assumption $\frac{\partial D_{mj}}{\partial C_{mj}} < 0$,
$C''_{mj} x_{mj}^2 + C'_{mj} x_{mj} \le 0$, the whole summation in the curly bracket can be bounded by
\[
|{\cal N}| \left( \sum_{(m,j) \in {\cal E}} \frac{\partial^2 D_{mj}}{\partial C_{mj}^2} (C'_{mj})^2 x_{mj}^2 +
\frac{\partial D_{mj}}{\partial C_{mj}} C''_{mj} x_{mj}^2 \right).
\]
Due to the peak power constraint, there exists a global upper bound on the achievable $SINR$ on all links, that is
$\bar x \ge \max_{(m,j) \in {\cal E}} x_{mj}$. Recall the definitions for $\kappa$, $\varphi$, $\bar B(D^0)$, and
$\underline B(D^0)$: $\kappa = \max_{0 \le x \le \bar x} (C'(x))^2 \cdot x^2 $, $\varphi = \min_{0 \le x \le \bar x}
C''(x) \cdot x^2$,
\[
\bar B(D^0) = \max_{(m,j) \in {\cal E}} \max_{D_{mj} \le D^0} \frac{\partial^2 D_{mj}}{\partial C_{mj}^2},
\]
and
\[
\underline B(D^0) = \min_{(m,j) \in {\cal E}} \min_{D_{mj} \le D^0} \frac{\partial D_{mj}}{\partial C_{mj}}.
\]
 We obtain
 \[
\bs v' \cdot H_{\bs\gamma} \cdot \bs v < \sum_{i \in {\cal N}} v_i^2 \bar S_i^2 |{\cal N}| |{\cal E}| \left[ \bar
B(D^0) \kappa + \underline B(D^0) \varphi \right],
 \]
i.e., $H_{\bs\gamma}$ is upper bounded by $|{\cal N}| |{\cal E}| \left[ \bar B(D^0) \kappa + \underline B(D^0) \varphi
\right]{\rm diag} \{ \left( \bar S_i^2  \right)_{i \in {\cal N}} \}$.  \qed

\subsection{Proof of Lemma~\ref{lma:MicroHessianBnd}}\label{app:MicroEtaHessian}

For convenience, we suppress the index $k$ and parameter $\lambda$.  Using the precise capacity formula
\eqref{eq:CapacityPrecise}, we derive the entries the Hessian matrix as
\[
\begin{array}{ll}
\vspace{1mm} \left[H_{\boldsymbol\eta_i}\right]_{jj} &= \displaystyle{\frac{\partial^2 D_{ij}}{\partial C_{ij}^2}
\left(\frac{G_{ij} P_i (K - 1)}{K G_{ij} P_{ij} + IN_{ij}} + \frac{ G_{ij} P_i}{IN_{ij}}\right)^2 + \frac{\partial
D_{ij}}{\partial C_{ij}} \left(\frac{- G_{ij}^2 P_i^2 (K-1)^2}{(K G_{ij} P_{ij} + IN_{ij})^2} + \frac{ G_{ij}^2
P_i^2}{IN_{ij}^2}\right)}
\\
\vspace{1mm} &< \displaystyle{\frac{\partial^2 D_{ij}}{\partial C_{ij}^2} \left[  \frac{ K G_{ij} P_i}{\sum_{m \ne i}
G_{mj} P_m + N_j}\right]^2 - \frac{\partial D_{ij}}{\partial C_{ij}} \left( \frac{ ({K-1})G_{ij} P_i}{\sum_{m \ne i}
G_{mj} P_m + N_j}\right)^2},
\\
\left[H_{\boldsymbol\eta_i}\right]_{lj} &= 0, \quad \forall l \ne j.
\end{array}
\]
Plugging the expressions \eqref{eq:barBij}-\eqref{eq:NRij}, we further bound the diagonal terms of
$H_{\boldsymbol\eta_i}$ as
\[
\left[H_{\boldsymbol\eta_i}\right]_{jj} < \left[\bar B_{ij}(D_i^k) K^2 - \underline B_{ij}(D_i^k)\left( {K-1} \right)^2
\right] NR_{ij}^2.
\]
Thus, the proof is complete. \qed

\bibliography{WirelessRouting}
\bibliographystyle{ieeetr}

\end{document}